\documentclass[12pt]{article}
\usepackage{graphicx}
\usepackage{amsmath}
\usepackage{amssymb}
\usepackage{bbm}

\usepackage{color}
\usepackage{psfrag}

\newcommand{\Tr}{\operatorname{Tr}}

\newcommand{\be}{\begin{equation}}
\newcommand{\ee}{\end{equation}}

\begin{document}

\begin{titlepage}

\begin{center}

\begin{flushright}
 CERN-TH-2003-119\\
 hep-lat/0306011\\[12ex]
\end{flushright}

\textbf{\large Observing string breaking with Wilson loops}
\\[6ex]

{Slavo Kratochvila$^{a,}$\footnote{skratoch@itp.phys.ethz.ch}
 and Philippe de Forcrand$^{ab,}$\footnote{forcrand@itp.phys.ethz.ch}}
\\[6ex]
{${}^a${\it Institute for Theoretical Physics, ETH Z\"{u}rich,
CH-8093 Z\"{u}rich, Switzerland}\\[1ex]
${}^b${\it CERN, Theory Division, CH-1211 Gen\`{e}ve 23, Switzerland}}
\\[10ex]
{\small \bf Abstract}\\[2ex]
\begin{minipage}{14cm}
{\small

An uncontroversial observation of adjoint string breaking is
proposed, while measuring the static potential from Wilson loops
only. The overlap of the Wilson loop with the broken-string state
is small, but non-vanishing, so that the broken-string groundstate can
be seen if the Wilson loop is long enough. We demonstrate this in
the context of the ${(2+1)d\;SU(2)}$ adjoint static potential,
using an improved version of the L\"uscher-Weisz exponential
variance reduction. To complete the picture we perform the more usual
multichannel analysis with two basis states, the unbroken-string
state and the broken-string state (two so-called gluelumps). \\

As by-products, we obtain the temperature-dependent static
potential measured from Polyakov loop correlations, and the
fundamental $SU(2)$ static potential with improved accuracy.
Comparing the latter with the adjoint potential, we see clear
deviations from Casimir scaling. }
\end{minipage}
\end{center}
\vspace{1cm}

\end{titlepage}

\section{Motivation}
\label{sec:motivation}

Quarks are linearly confined inside hadrons by a force called the strong interaction.
Therefore, we cannot see single quarks: This is the basis of the stability of the matter
we are formed of. One can study this force by analysing the energy between a static
colour charge and a static anticharge. Unlike in the
case of the electromagnetic force, this energy is, as a consequence of linear
confinement, squeezed into a long flux tube.
This flux tube is a string-like object. Therefore, one can ask whether this string actually breaks
when it reaches a certain length.
This breaking of the string corresponds to the screening of the static charges by a
virtual matter-antimatter pair created from that very energy stored in the string.
The energy of the groundstate of the system, the so-called static potential,
completely changes its qualitative behaviour as a function of the distance between the two static
charges and can
therefore be used to detect string breaking.
There are two main situations where string breaking can be studied:
$(i)$ when one deals with static charges in the fundamental representation,
which can only be screened by other fundamental particles, such as dynamical quarks or
fundamental Higgs fields; $(ii)$ when one considers static charges in the adjoint representation
which can be screened by the gluons of the gauge field.
To avoid the simulation of costly dynamical quarks or of Higgs fields,
we simply consider here adjoint static
charges. The bound state of a gluon and an adjoint static charge is called a ``gluelump''.
Therefore, the breaking of the adjoint string leads to the creation of two of these
gluelumps.

Three approaches have been used to measure the static
potential and study string breaking:
\begin{itemize}
  \item Correlation of Polyakov loops, at finite temperature
\cite{Laermann:1998gm}.
  \item Multichannel Ansatz (also known as Variational Ansatz) using two types of
operators: one for the string-like state and one for the broken-string state
\cite{Michael:1991nc,Stephenson:1999kh}.
  \item Wilson loops \cite{Born:1994,Poulis:1995nn}.
\end{itemize}

String breaking has been seen using the first two methods, but no
clear signal has been observed using the third one. The failure of
the Wilson loop method seems to be due to the poor overlap of the
Wilson loop operator with the broken-string state. It has even
been speculated that this overlap is exactly zero
\cite{Gliozzi:1999cv}. This is why we have a closer look at this
problem, taking advantage of recent, improved techniques to
measure long Wilson loops.
Preliminary results have been presented in \cite{Kratochvila:2002vm}.\\

In the next Section we recall notions about the static potential
and its relation to the Wilson loop. In Section
\ref{sec:approaches}, we take the three methods into more detailed
consideration, to be able to discuss our results in all
approaches. We explain the difficulties of measuring string
breaking using Wilson loops only. In Section \ref{sec:method}, we
present the techniques we applied, such as adjoint smearing and
improved exponential error reduction for Wilson loops. We also
discuss the methods used to analyse our data. For all three
methods, results are shown in Section \ref{sec:results}, followed
by the conclusion.

\section{Static potential}
\label{sec:staticpotential}

We consider a pure $SU(2)$ gauge system with a static charge and a
static anticharge separated by a distance $R$. Although we mainly focus
on measuring the static potential using Wilson loops only, we also consider the
Multichannel Ansatz. Therefore, here we discuss the issue of measuring the groundstate
and excited-state energies in a more general way.

The Hilbert space
of the Hamiltonian is spanned by its orthonormal eigenbasis
$\Psi^{(n)}(R)$. The corresponding energies are $E_0 < E_1 < ...$,
where $E_0$, the groundstate energy, is called the static
potential. If we knew these eigenstates, we could extract the
energies by measuring their time evolution: \be
\label{eq:staticpotential_gamma_matrix}
    \Gamma^{(n)}(R,T) = \langle \Psi^{(n)}(R) \mid {\bf \hat{T}}^T \mid \Psi^{(n)}(R) \rangle = e^{-E_n(R) T}
\ee where ${\bf \hat{T}}$ is the transfer operator,  ${\bf
\hat{T}} = \sum_{n=0}^{\infty} |\Psi^{(n)}(R) \rangle  e^{-E_n(R)}
\langle \Psi^{(n)}(R)|$. Since we do not know these
eigenstates explicitly, let us consider an arbitrary linear combination
$\Phi(R)$. We can expand this state in the eigenbasis \be
    |\Phi(R) \rangle =  \sum_n \; \langle \Psi^{(n)}(R) | \Phi(R) \rangle \; |\Psi^{(n)}(R)\rangle \;.
\ee
The temporal evolution is given by
\be \label{eq:staticpotential_temp_evolution}
   \langle \Phi(R) \mid {\bf \hat{T}}^T \mid \Phi(R) \rangle =  \sum_n \;
   | \langle \Phi(R) \mid \Psi^{(n)}(R)\rangle |^2 \; e^{-E_n(R) T} \equiv \sum_n \; c_n  e^{-E_n(R) T}   \;.
\ee
Let us introduce a finite set of states $\phi_i(R)$ which we know how to measure. We set
\be
    \Phi(R) = \sum_i a_i(R) \phi_i(R) \;.
\ee
To be explicit, let us expand Eq.(\ref{eq:staticpotential_temp_evolution}) and define the correlation matrix
$V_{ij}(R,T)$
\be \label{eq:staticpotential_correlation_matrix}
    V_{ij}(R,T) = \langle \phi_j(R) \mid {\bf \hat{T}}^T \mid \phi_i(R) \rangle \;.
\ee
so that
\begin{align}
    \langle \Phi(R) \mid {\bf \hat{T}}^T \mid \Phi(R) \rangle
    =  \sum_{i,j} a_j(R) a_i(R) V_{ij}(R,T) \;.
\end{align}

A lemma, which is proven in \cite{Luscher:1990}, states for the eigenvalues $\lambda^{(n)}(R,T)$
of the correlation matrix $V_{ij}$
\be \label{eq:staticpotential_diagonalisation_eigenvalues}
    \lambda^{(n)}(R,T) \underset{T \rightarrow \infty}{=} f^{(n)}(R) e^{-E_n T} \left( 1 + O(e^{-T \Delta E_n})\right)
\ee
where $f^{(n)}(R)>0$ and $\Delta E_n = \underset{m \neq n}{\min} | E_n - E_m |$.
In general, the correction term cannot be neglected and it will play an essential
role in our study. The way to determine
the eigenvalues $\lambda^{(n)}(R,T)$ and estimate the energies $E_n$ is
described in Subsection \ref{subsec:method_diagonalisation_procedure}.\\

The finite set of states $\phi_i(R)$ is created by applying some operators on the vacuum.
These states are chosen to model the expected ones, the unbroken-string and the broken-string state.
We can build a string-like state $\phi_s(R)$  by a spatial line $S_{s}(R)$ of links of length $R$
 where $s$ denotes the number of spatial smearing iterations
(see Subsection \ref{subsec:method_adjointsmearing}),
or a broken-string state $\phi_G(R)$ of length $R$,
using $G(R)$, a "clover" discretisation of $F_{\mu\nu}$ around the two static charges
(see Subsection \ref{subsec:method_gluelumps}).
The correlation matrix is then
\begin{align}
    V_{ij}(R,T)  &=  \langle \phi_j(R) \mid {\bf \hat{T}}^T \mid \phi_i(R) \rangle \; , \; i,j=s, G \\
                  &=  \left(
    \begin{matrix}
        \includegraphics{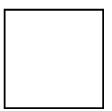} & \includegraphics{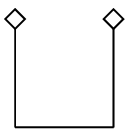} \\
        \includegraphics{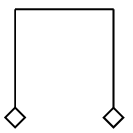} & \includegraphics{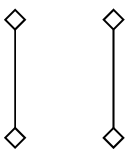}
    \end{matrix}
    \right)
    =
    \left(
    \begin{matrix}
        S_s S_s(R,T) & S_s G(R,T) \\
        G S_s(R,T) & GG(R,T)
    \end{matrix}
    \right)
     \label{eq:staticpotential_multichannel_ansatz}
\end{align}

\begin{figure}[!t]
  \centering
    \begin{minipage}{6.00cm}
    \includegraphics[width=6.0cm]{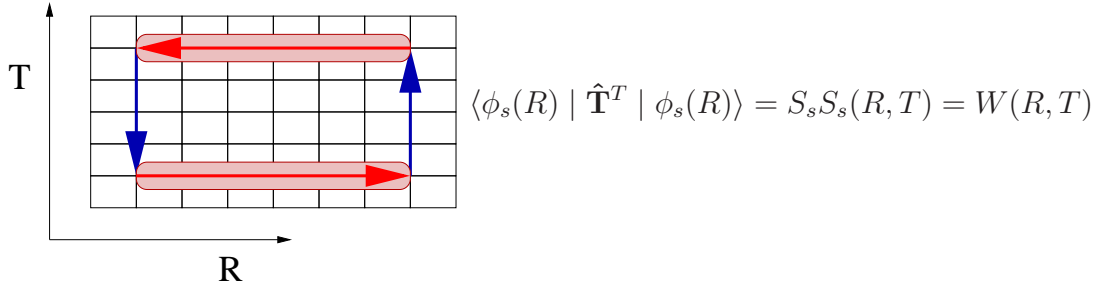}
    \end{minipage}
    \begin{minipage}{6.00cm}
    \begin{center}
    \begin{equation*}
    \label{eq:staticpotential_wilson_loop} \langle \phi_s(R)
    \mid {\bf \hat{T}}^T \mid  \phi_s(R) \rangle = S_s S_s(R,T) =
    W(R,T)
    \end{equation*}
    \vspace{0.75cm}
    \end{center}
    \end{minipage}
    \caption{The Wilson loop channel. Information about the groundstate energy, the so-called static potential,
    can be extracted. }
    \label{fig:staticpotential_wilson_loop}
\end{figure}

Using the diagonalisation procedure\footnote{Throughout this paper, we call
the \emph{multichannel Ansatz} the approach to
obtain eigen-energies and -states using a multichannel approach, i.e.
measuring correlations between states
created by a finite set of operators. We call the \emph{diagonalisation procedure}
the numerical procedure we use to extract
information about eigen-energies and -states from a given correlation matrix of
type Eq.(\ref{eq:staticpotential_correlation_matrix}). }
 (see Subsection
\ref{subsec:method_diagonalisation_procedure}) one can reconstruct
$\Gamma^{(n)}(R,T)$ (Eq.(\ref{eq:staticpotential_gamma_matrix})) for small $n$, the
lowest energies, hence the static potential, and the
overlaps of the string-like state and the broken-string state with
the corresponding eigenstates. In previous studies
\cite{Michael:1991nc,Stephenson:1999kh}, this correlation matrix
has been used in the multichannel Ansatz to show string breaking.
We will confirm these results. But since the multichannel Ansatz
has been objected to (see next Section), we will also show that the full
information about the static potential can be extracted using only
the $GG$-channel or only the $S_s S_s(R,T)$ channel (the Wilson loop $W(R,T)$
(see Fig.~\ref{fig:staticpotential_wilson_loop})).
Using the above notation, this corresponds to setting $a_s=1$, the other $a_i=0$.
Eq.(\ref{eq:staticpotential_temp_evolution}) is now given by
\begin{align} \label{eq:staticpotential_expansionwilsonloop}
   \langle \phi_s(R) \mid {\bf \hat{T}}^T \mid \phi_s(R) \rangle & =  S_s S_s(R,T) = W(R,T) \notag \\
   & = \sum_n \;
   | \langle \phi_s(R) \mid \Psi^{(n)}(R)\rangle |^2 \; e^{-E_n(R) T} = \sum_n \; c_n e^{-E_n(R) T}
\end{align}

We can truncate this sum of exponentials at $n=l$ if all the states we neglect are strongly suppressed.
For all $k>l$, we demand
\be
    1 \gg \frac{c_{k}(R)}{c_{l}(R)} e^{-(E_k(R)-E_l(R)) T_{\textnormal{min}}(R)}
\ee
which defines a $T_{\textnormal{min}}(R)$ implicitly. In particular,
we consider the following two-mass Ansatz ($l=1$)
\be \label{eq:staticpotential_two_mass_ansatz}
  W(R,T) = c_0 e^{-V(R) T } + c_1 e^{-E_1(R) T }\;,\;T > T_{\textnormal{min}}
\ee
where $c_0$ and $c_1$ are the overlaps of the state $| \phi_s \rangle$ with the groundstate, which
is the static potential, and the first
excited state respectively.

\section{The three approaches}
\label{sec:approaches}

The adjoint Wilson loop shows a good overlap with
the unbroken-string but not with the broken-string state. This is natural: The Wilson loop observable creates
a static quark-antiquark pair together with a flux tube joining them. Therefore, the broken-string state,
which consists of two isolated gluelumps, has poor overlap with
the flux-tube state, hence with the Wilson loop.
For $R<R_b$ the unbroken-string state is
the groundstate, therefore in Eq.(\ref{eq:staticpotential_two_mass_ansatz}) $c_0$ is large compared to $c_1$.
In this regime, additionally $E_1 > V$ by definition, therefore the second exponential
in Eq.(\ref{eq:staticpotential_two_mass_ansatz}) is negligible
for $T>T_{\textnormal{min}}$. An important issue is what happens for $R$ larger than $R_b$.
The broken-string state becomes the groundstate. Its energy  $V(R)$ is smaller than the
unbroken-string state energy $E_1(R)$ (level crossing has occurred).
But for small $T$, the unbroken-string state still dominates over the broken-string state
because $c_0 \ll c_1$. Of course, this domination holds only up
to a temporal extent $T = T_P$, where $T_P$ is the turning point defined by the
equality of both terms on the right-hand side of (\ref{eq:staticpotential_two_mass_ansatz}).
For $T>T_P$, the broken-string groundstate becomes "visible" in the exponential decay of
Eq.(\ref{eq:staticpotential_two_mass_ansatz}).
The value of this
$T_P$ is crucial to be able to detect string breaking using Wilson loops only.
\be \label{eq:approaches_turningpoint}
    T_P = \frac{\ln\frac{c1}{c0}}{E_1(R)-V(R)} \;.
\ee

Based on the heavy quark expansion, the strong coupling model of Ref.~\cite{Drummond:1998ar}
leads to estimates of the ratio $\frac{c_1}{c_0}$ and of $T_P$. As an example we consider
the distance $R=12a = 1.228(1)$ fm,
whereas string breaking occurs at $R_b \approx 10a = 1.023(1)$ fm
(The lattice spacing $a$ is calculated in the beginning of Section \ref{sec:results}).
The energy of the broken-string state is $V(R) \approx 2 \times M(Qg)$, where $M(Qg)=1.03(2)a^{-1}$
is the mass of a gluelump (measured independently in Subsection \ref{subsec:method_gluelumps}).
Within this model, the ratio $\frac{c_1}{c_0}$ is around
$\frac{c_1}{c_0} \sim e^{M(Qg) R} \underset{R = 12a}{\approx} 2\times 10^5$. Therefore,
the turning point is estimated to be
\be \label{eq:approaches_turningpoint_estimate}
T_P^{(est)} = \frac{M(Qg) R}{E_1(R) - 2M(Qg)} \underset{R = 12a}{\approx} 42a
\ee
using
the value $E_1(R=12a)= 2.34(1) a^{-1}$ of our results in advance. Wilson loops of
size $12a \times 42a$, i.e. about $1.2$fm $\times 4.3$fm, or larger are needed to observe
string breaking according to this model.

Based on a topological argument, Ref.~\cite{Gliozzi:1999cv} even suggests that there may be no
overlap at all: $c_0=0$ for $R>R_b$. Adding matter fields gives rise to the formation of holes
in the world sheet of the Wilson loop, reflecting pair creation. The average hole size
leads to two different phases of the world sheet. In the normal phase,
holes are small and the Wilson loop still fulfills an area law, $W(R,T) \sim e^{-\sigma R T}$,
where $\sigma$ is the string tension renormalised by the small holes.
This phase corresponds to the unbroken-string case and
a screening of the static charges cannot be observed. The other phase is called the
tearing phase, where holes of arbitrarily large size can be formed.
As a consequence, the Wilson loop follows a perimeter law, $W(R,T) \sim e^{-c T}$.
This corresponds to the broken-string state, since the groundstate energy remains constant.
\cite{Gliozzi:1999cv} speculates that the Wilson loop is in the normal phase, and analyticity
prevents it from changing phase, so that string breaking cannot be seen.
This would explain why, for instance, the authors of \cite{Poulis:1995nn} could not observe
string breaking even at a distance $R \approx 2R_b$. Note however, that their temporal extent
was $T \leq 3$.

A simple argument gives a necessary condition for the observation of
string breaking if we only use ordinary (non-smeared) Wilson loops $W(R,T)$.
If $R > R_b$, where $R_b$ is the string breaking distance,
but $T < R_b$, we can relabel the $R$ and $T$ directions, so that now $R < R_b$ and $T > R_b$.
In that case, string breaking is not visible since the new spatial extent is $R < R_b$,
and the Wilson loop must still fulfill the area law. Therefore, \emph{both} sides $R$ and $T$
should be larger than $R_b$. Replacing ordinary Wilson loops by spatially smeared ones may
relax this requirement somewhat. \\

As already mentioned in the previous Section, an improved determination of the
static potential can be achieved by a variational
superposition of the unbroken-string and the broken-string states. In other words,
the multichannel Ansatz enlarges the operator space to a multichannel approach.
The unbroken-string state is realised via the flux tube of the Wilson loop,
the broken-string state is modelled by
considering two gluelumps - separated static charges surrounded by gluons which
screen the "interior" colour charge. We end up with the two-channel transfer matrix
$V_{ij}$ of Eq.(\ref{eq:staticpotential_multichannel_ansatz}).

Nevertheless, this method has been criticised \cite{Kallio:2000jc}.
One may claim that string breaking is \emph{built into} the multichannel Ansatz due to the
explicit inclusion of both states. Moreover,
the behaviour in the continuum limit ($\beta \rightarrow
\infty$) must be considered. If the off-diagonal element
$S_s G(R,T) = G S_s(R,T)$ is zero in this limit,
string breaking does not actually happen: the Wilson loop does not communicate
with the broken-string state; the eigenvalues merely cross each other at $R=R_b$.
It is only if the off-diagonal elements are different from zero, that the eigenstates are a mixture
of both states. A small overlap at different $\beta$-values has indeed been confirmed
by the mixing analysis in the multichannel Ansatz, as can be seen
in Refs. \cite{Michael:1991nc,Stephenson:1999kh},
which deal with this issue.
However, whether this overlap vanishes or not
for $a \rightarrow 0$ still has to be checked in detail.
Here, we do not address this question of continuum
extrapolation due to technical difficulties: The method we use is more efficient at smaller
$\beta$. Therefore we consider only one $\beta$-value, which we choose as small as
possible while staying in the scaling region.\\

The Wilson loop method and the multichannel Ansatz work at zero temperature,
where the question to be answered is: What is the
groundstate of a system with two static charges? In the
context of the Polyakov loop method we have contributions of
temperature-dependent effects, and the question to be answered is different:
What is the free energy of a system with two static charges coupled to a
heatbath? Nevertheless, it is an interesting issue and, as a by-product,
we can also measure the correlations of Polyakov
loops, which are in our case of adjoint charges
$P({\bf x}) =\frac{1}{3} \Tr \prod_{i=0}^{N_t-1} U_{\textnormal{adj}}^{(t)} ({\bf x}, i)$.
This results in a temperature-dependent potential $V_{\cal{T}}(R)$ \cite{McLerran:1981}:
\be \label{eq:approaches_polyakov}
    \langle P(0) P^*(R) \rangle = e^{-V_{\cal{T}}(R)/\cal{T}}
\ee

The flattening of the potential $V_{\cal{T}}(R)$ at large $R$
has been seen in QCD with dynamical fermions,
although in practice only at temperatures close to or
above the critical deconfinement temperature. Unlike the multichannel Ansatz,
this method builds in no prejudices about the structure of the
 groundstate wave function.

\section{Technical details}
\label{sec:method}

We are considering Wilson loops in the adjoint representation (for a definition see the following
Subsection). The choice of the representation has a direct impact on the static
potential, which is at lowest order in perturbation theory in (2+1) dimensions \cite{Schroder:1999sg}
\be \label{eq:method_pert_staticpotential}
    V_P(R) \sim - C_2 \frac{g_0^2}{2\pi} \ln R \Lambda + O(g_0^4/\Lambda)\;.
\ee
$C_2$ is the value of the quadratic Casimir operator of the representation of the static charges, i.e.
\begin{itemize}
    \item fundamental representation: $C_2(F)\; \mathbbm1_{2\textnormal{x}2}= \frac{3}{4}\; \mathbbm1_{2\textnormal{x}2}$
    \item adjoint representation: $C_2(A)\; \mathbbm1_{3\textnormal{x}3}= 2\; \mathbbm1_{3\textnormal{x}3}$
\end{itemize}

The important point is that, in the regime of perturbation theory, i.e. at small distances $R$,
at lowest order, the adjoint static potential
$V_{\textnormal{adj}}(R)$ differs from the fundamental static potential
$V_{\textnormal{fund}}(R)$ by a factor $\frac{8}{3}$. Assuming, for simplicity, that the
ratio remains the same at larger $R$ (this issue is discussed in Subsection \ref{subsec:results_casimir}),
the adjoint potential is much more difficult to measure than the fundamental one:
 The Wilson loop is
$W(R,T) \sim e^{-V(R) T}$, therefore the signal decreases much faster with $R$ or $T$ in the adjoint
representation.
This is the price to pay if we consider adjoint static charges instead of fundamental static
charges in order to avoid the simulation of dynamical quarks. We need a sophisticated method
of exponential error reduction (see Subsection \ref{subsec:method_exponentialerrorreduction})
to detect very small signals: The magnitude of each measured Wilson loop is $W_i(R,T) \sim O(1)$ while the
average to be detected, as it will turn out, is $W(R,T)\sim O(10^{-40})$. Using ordinary methods, $10^{80}$ measurements
would be needed.

\subsection{Adjoint smearing}
\label{subsec:method_adjointsmearing}

It is very desirable to reduce contributions from excited states $\Psi^{(n \neq 0)}$ to the
Wilson loop $W(R,T)$: The turning point Euclidean time $T_P$ (Eq.(\ref{eq:approaches_turningpoint}))
is reduced, and the accuracy on the groundstate potential is greatly improved. To this end,
we smear adjoint links spatially.
In $SU(2)$, a matrix $U_{\textnormal{fund}}$ in the fundamental representation can be mapped onto a
$3 \times 3$ real link matrix $U_{\textnormal{adj}}$ in the adjoint representation by
\begin{equation}\label{eq:method_adjointlink}
U_{\textnormal{adj}}^{\alpha\beta}(U_{\textnormal{fund}})=\frac{1}{2} \sigma_{li}^\alpha
U_{\textnormal{fund},ij} \sigma_{jk}^\beta U_{\textnormal{fund},kl}^\dagger \;.
\end{equation}
where the $\sigma^\alpha$ are the Pauli matrices; $\alpha,\beta=1..3$; $i,j,k,l=1..2$.\\
The smearing can be done by setting the new adjoint link as the $SO(3)$ projection of the old link
plus a weighted sum of the spatial staples:
\be \label{eq:method_smearing}
U_{\textnormal{adj}}'(x) = \textnormal{Proj}_{SO(3)} \left( U_{\textnormal{adj}}(x) +
\alpha \sum_{i=1}^4 \textnormal{Adjoint Spatial Staple}_i \right)
\ee
where we choose $\alpha=0.5$. We consider three different smearing levels: 15, 30 and
60 iterations of Eq.(\ref{eq:method_smearing}). For details
and usage, see Subsection \ref{subsec:method_improved_spatial_transporter}.
We define our projection of an arbitrary matrix $\tilde{A}$ onto $A \in SO(3)$ by maximising
$\Tr \tilde{A}^\dagger A$. This can be performed using the singular value decomposition:
Every $M \times  N$ matrix $\tilde{A}$ ($M \geq N$) can be written as the product of
a column-orthogonal $M \times  N$ matrix $U$,
a diagonal $N \times N$ matrix $\Delta$ with positive or zero elements (the singular values),
and an orthogonal $N \times  N$ matrix $V^\dagger$.
\be
    \tilde{A} = U \Delta V^\dagger
\ee
Since in our case $M=N=3$, both $U$ and $V$ are elements of $SO(3)$, and we get the projection
of $\tilde{A}$ onto $SO(3)$ by
\be
    A = \textnormal{Proj}_{SO(3)} \left(  \tilde{A} \right) = U V^\dagger
\ee

For completeness we also describe the fundamental smearing procedure used in Subsection \ref{subsec:method_gluelumps}.
Instead of considering adjoint links,
we use the same method but in the fundamental representation:
\be \label{eq:method_smearing_fundamental}
U'(x) = \textnormal{Proj}_{SU(2)} \left( U(x) + \alpha \sum_{i=1}^4 \textnormal{Spatial Staple}_i \right)
\ee
where we take $\alpha=0.5$ and project back onto $SU(2)$ by
\be
    U'(x) = \textnormal{Proj}_{SU(2)} \left(  \tilde{U}'(x) \right)=  \frac{\tilde{U}'(x)}{\sqrt{\det \tilde{U}'(x)}}
\ee

\subsection{Exponential error reduction}
\label{subsec:method_exponentialerrorreduction}

An adjoint $R$ by $T$ Wilson loop in the $(x,t)$-plane consists of
two adjoint spatial transporters of length $R$, which
we call $\textbf{L}(0)$ and $\textbf{L}(T)^\dagger$,
and two temporal sides of length $T$, which we write explicitly as
$\textbf{U}(0)^\dagger=\prod_{i=0}^{T-1}  U_{\textnormal{adj}}^{\dagger(t)}(x,y,t+i)$ and
${\textbf{U}(R)=\prod_{j=T-1}^{0} U_{\textnormal{adj}}^{(t)}(x+R,y,t+j)}$\footnote{To make clearer
the distinction between the two temporal sides of the Wilson loop, we use separate indices $i$
and $j$ for the two running time-coordinates.}.
The expectation value of the Wilson loop can be written as
\be \label{eq:method_wilson_loop_exp}
 W(R,T) = \frac{1}{Z} \int [DU] \Tr \{ \textbf{U}(0)^\dagger \textbf{L}(T)^\dagger \textbf{U}(R) \textbf{L}(0)\}
  \; e^{-S[U]}
\ee

An exponential error reduction is possible because of the locality of the action, which in
our case is the Wilson plaquette action. The main idea is to write the average of a product
as a product of averages.

\subsubsection{Multihit-method}

One possibility to reduce the variance of the Wilson loop
observable, is to reduce the variance of a single link
contribution. The Multihit method \cite{Parisi:1983} takes the
average of many samples of one particular link with all other
links held fixed. As we will show now, all the temporal links\footnote{The situation is
different considering spatial links: Since we have
smeared them spatially, we cannot "hold all the other
links fixed".}
$U_{\textnormal{adj},k}^{(t)}$ in Eq.(\ref{eq:method_wilson_loop_exp}) can be treated like this for
Wilson loops with a spatial extent $R \geq 2$.
 In the first step we split the action as $S = S'[U'] +
\sum_k S_k''[U_k]$. $S_{k}''[U_{k}]$
is the local part of the action that contains the fundamental link
$U_k$  corresponding to $U_{\textnormal{adj},k}$.
The Multihit-method can be applied also to $U_{\textnormal{adj},l}$ if $S_k''[U_k]$ does
not depend on $U_l$ for $k \neq l$. Therefore, since we use
the Wilson plaquette action, this condition is satisfied if $R \geq 2$.
We can then apply the Multihit-method on all time-like links
 and Eq.(\ref{eq:method_wilson_loop_exp})
can be rewritten as
\begin{align}
 \frac{1}{Z}\int [DU'] \Tr  & \left( \prod_{i=0}^{T-1} \int dU_i U^\dagger_{\textnormal{adj},i}  \; e^{-S_i''[U_i]}\right)
    \textbf{L}(T)^\dagger \times \\
    & \left( \prod_{j=T-1}^0 \int dU_j U_{\textnormal{adj},j}  \; e^{-S_j''[U_j]}\right)
    \textbf{L}(0) \; e^{-S'[U']} = \notag
\end{align}
\begin{equation}\label{eq:method_wilson_loop_exp_multi}
 \frac{1}{Z} \int [DU] \Tr \left(\prod_{i=0}^{T-1} \bar{U}^\dagger_{\textnormal{adj},i} \right)
    \textbf{L}(T)^\dagger \left( \prod_{j=T-1}^0 \bar{U}_{\textnormal{adj},j} \right)
    \textbf{L}(0)
    \; e^{-S[U]}
\end{equation}
where the Multihit-average is given by the one-link integral
\be
    \bar{U}_{\textnormal{adj},i} = \frac{\int dU_i U_{\textnormal{adj},i}  \; e^{-S_i''[U_i]}}
    {\int dU_i e^{-S_i''[U_i]}}
\ee
In simple cases, as  in pure $SU(2)$,
the Multihit-average can even be calculated analytically. Namely,
$S_i''[U_i]=-\beta \frac{1}{2} \Tr U_i W^\dagger$,
where $W$ is the sum of the four (in $3d$) fundamental staples, and
\be
    \bar{U}_{\textnormal{adj},i} = \hat{W}_{\textnormal{adj}} \frac{I_3(\beta w)}{I_1(\beta w)}
\ee
where $w=\sqrt{\det W}$, $\hat{W} = W/w$
is the projection of the staple-sum
onto $SU(2)$ and $\hat{W}_{\textnormal{adj}}$ represents $\hat{W}$ in the adjoint
representation via Eq.(\ref{eq:method_smearing}). \\

Since the variance of each time-like link entering $W(R,T)$ is reduced, the variance reduction
in $W(R,T)$ is exponential in $T$. The coefficients have been estimated in \cite{Michael:1985}.
For fundamental Wilson loops, the reduction is about $(0.8^2)^T = e^{-0.45T}$, and for adjoint
loops about $(0.5^2)^T = e^{-1.39 T}$.

\subsubsection{Multilevel-method}

Although the Multihit-method was revolutionary in 1983, the error reduction
was not strong enough to enlarge measurable Wilson loops to temporal
extents $T$ sufficient to be able to observe string breaking.
In Section \ref{sec:approaches}, we suggested a heuristic argument,
that $T$ should be at least as large as the string breaking distance, which in our case
is at $R_b \sim 10a$. The heavy quark
expansion even results in an estimation of $T_P^{(est)} \approx 42a$
(see Eq.(\ref{eq:approaches_turningpoint_estimate})). \\

M.~L\"uscher and P.~Weisz generalise the Multihit method
from single time-like links to link-link correlators $\textbf{T}(R,t'=n a)$ \cite{Luscher:2001up}.
Using our notation from above,
\be
    \textbf{T}(R,t'=n a)_{\alpha\beta\gamma\delta} =
        U_{\textnormal{adj}}^{*(t)}(x,y,t+i=n a)_{\alpha\beta} \; U_{\textnormal{adj}}^{(t)}(x+R,y,t+j=n a)_{\gamma\delta} \;.
\ee
A single Wilson loop $W_s(R,T)$ can be written, using the tensor multiplication defined by
\be \label{eq:method_tensor_multiplication}
    \{ \textbf{T}(R,n a)\textbf{T}(R,(n+1) a) \}_{\alpha\beta\gamma\delta}
      =
        \textbf{T}(R,n a)_{\alpha\lambda\gamma\epsilon}
        \textbf{T}(R,(n+1) a)_{\lambda\beta\epsilon\delta}
\ee
as
\be
    W_s(R,T) = \textbf{L}(0)_{\alpha\gamma}
    \{\textbf{T}(R,0)\textbf{T}(R,1a)...\textbf{T}(R,(T-1)a)\}_{\alpha\beta\gamma\delta}
    \textbf{L}(T)^*_{\beta\delta} \;.
\ee

Just like in the Multihit-method where we considered the average links $\bar{U}_i$,
here time-slice expectation values of a link-link-correlator $\textbf{T}(R,n a)$, denoted by $[\;.\;]$,
can be obtained by sampling over the corresponding
sublattice which is in this case the time-slice at time $t'=n a$.
This sublattice can be studied independently of the surrounding lattice provided the spatial link
variables at the boundaries are held fixed.
This is a consequence of the time-locality of the gauge action. Using a self-explanatory notation:
\begin{equation}
    \textcolor{blue}{[}\textbf{T}(R,n a)\textcolor{blue}{]} \equiv \frac{1}{Z_{\textnormal{sub}}}
    \int [DU]_{\textnormal{sub}} \textbf{T}(R,n a) e^{-S[U]_{\textnormal{sub}}} \;,
\end{equation}
the expectation value of the Wilson loop can be written in the form
\be \label{eq:method_wilson_loop_exp_lw}
    W(R,T) = \langle \textbf{L}(0)_{\alpha\gamma}
    \{\textcolor{blue}{[}\textbf{T}(R,0)\textcolor{blue}{][}\textbf{T}(R,1a)\textcolor{blue}{]}...
    \textcolor{blue}{[}\textbf{T}(R,(T-1)a)\textcolor{blue}{]}\}_{\alpha\beta\gamma\delta}
    \textbf{L}(T)^*_{\beta\delta} \rangle \;.
\ee

The restriction of fixed spatial links at the boundaries becomes manifest in the fact that only
the \emph{temporal} links on the time-slice at time $t'=n a$ are allowed to be updated when evaluating $[\;.\;]$
with Monte Carlo methods. For our project, this is a severe obstacle which limits the
efficiency of the exponential error reduction. It can be circumvented by using a hierarchical scheme based
on identities like
\begin{gather}
    \textcolor{blue}{[}\textbf{T}(R,n a)\textcolor{blue}{][}\textbf{T}(R,(n+1) a)\textcolor{blue}{]}
     = \notag \\
    \textcolor{red}{\Bigl[}\textbf{T}(R,n a)\textbf{T}(R,(n+1) a)\textcolor{red}{\Bigr]} =
    \textcolor{red}{\Bigl[}\textcolor{blue}{[}\textbf{T}(R,n a)
    \textcolor{blue}{][}\textbf{T}(R,(n+1) a)\textcolor{blue}{]}\textcolor{red}{\Bigr]}
    \label{eq:method_hierarchical_id}
\end{gather}

\begin{figure}[!t]
  \centering
\psfrag{[[[T(0)][T(a)]][[T(2a)][T(3a)]]]}{\tiny
\textcolor{green}{\hspace{-0.25cm}$\biggl[$}
\textcolor{red}{\hspace{-0.1cm}$\Bigl[$}
\textcolor{blue}{\hspace{-0.075cm}$[$}
\textcolor{black}{\hspace{-0.075cm}$\textbf{T}(0)$}
\textcolor{blue}{\hspace{-0.075cm}$][$}
\textcolor{black}{\hspace{-0.075cm}$\textbf{T}(a)$}
\textcolor{blue}{\hspace{-0.075cm}$]$}
\textcolor{red}{\hspace{-0.1cm}$\Bigr]$}
\textcolor{red}{\hspace{-0.1cm}$\Bigl[$}
\textcolor{blue}{\hspace{-0.075cm}$[$}
\textcolor{black}{\hspace{-0.075cm}$\textbf{T}(2a)$}
\textcolor{blue}{\hspace{-0.075cm}$][$}
\textcolor{black}{\hspace{-0.075cm}$\textbf{T}(3a)$}
\textcolor{blue}{\hspace{-0.075cm}$]$}
\textcolor{red}{\hspace{-0.1cm}$\Bigr]$}
\textcolor{green}{\hspace{-0.125cm}$\biggr]$}
}
\psfrag{[[T(0)][T(a)]]}{\tiny
\textcolor{red}{\hspace{-0.1cm}$\Bigl[$}
\textcolor{blue}{\hspace{-0.075cm}$[$}
\textcolor{black}{\hspace{-0.075cm}$\textbf{T}(0)$}
\textcolor{blue}{\hspace{-0.075cm}$][$}
\textcolor{black}{\hspace{-0.075cm}$\textbf{T}(a)$}
\textcolor{blue}{\hspace{-0.075cm}$]$}
\textcolor{red}{\hspace{-0.1cm}$\Bigr]$}
}
\psfrag{[[T(2a)][T(3a)]]}{\tiny
\textcolor{red}{\hspace{-0.1cm}$\Bigl[$}
\textcolor{blue}{\hspace{-0.075cm}$[$}
\textcolor{black}{\hspace{-0.075cm}$\textbf{T}(2a)$}
\textcolor{blue}{\hspace{-0.075cm}$][$}
\textcolor{black}{\hspace{-0.075cm}$\textbf{T}(3a)$}
\textcolor{blue}{\hspace{-0.075cm}$]$}
\textcolor{red}{\hspace{-0.1cm}$\Bigr]$}
}
\psfrag{[T(3a)]}{\tiny
\textcolor{blue}{\hspace{-0.175cm}$[$}
\textcolor{black}{\hspace{-0.075cm}$\textbf{T}(3a)$}
\textcolor{blue}{\hspace{-0.075cm}$]$}
}
\psfrag{[T(2a)]}{\tiny
\textcolor{blue}{\hspace{-0.175cm}$[$}
\textcolor{black}{\hspace{-0.075cm}$\textbf{T}(2a)$}
\textcolor{blue}{\hspace{-0.075cm}$]$}
}
\psfrag{[T(a)]}{\tiny
\textcolor{blue}{\hspace{-0.175cm}$[$}
\textcolor{black}{\hspace{-0.075cm}$\textbf{T}(a)$}
\textcolor{blue}{\hspace{-0.075cm}$]$}
}
\psfrag{[T(0)]}{\tiny
\textcolor{blue}{\hspace{-0.175cm}$[$}
\textcolor{black}{\hspace{-0.075cm}$\textbf{T}(0)$}
\textcolor{blue}{\hspace{-0.075cm}$]$}
}
  \mbox{\includegraphics[width=12.0cm]{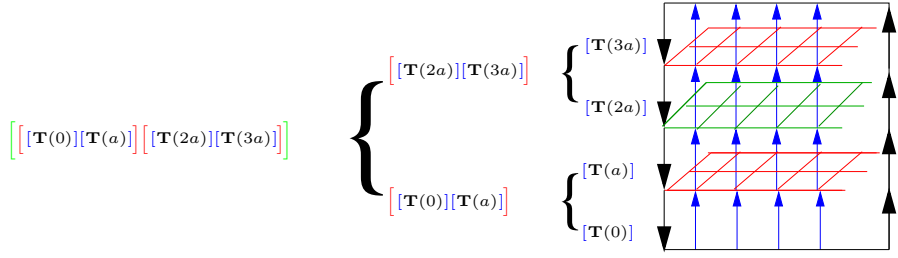}}
  \caption{Hierarchical scheme. Using the three-level method as described in the text decreases
  the statistical error exponentially. While a one-level approach only allows
  to sample over the temporal links, a multi-level approach also makes it possible
  to update the interior spatial links, which improves the error reduction.}
  \label{fig:method_hierarchical}
\end{figure}

The impact on the Monte Carlo method is, that we are also allowed to sample over spatial links
on the time-slice $(n+1) a$ since the boundary of the so-called second-level average
$\textcolor{red}{\Bigl[}\;.\;.\;\textcolor{red}{\Bigr]}$ now
consists of the spatial links in the time-slices $n a$ and $(n+2) a$.
The possibility to use this two-level scheme allows us to measure
long Wilson loops almost up to the desired accuracy.
We actually implement the following three-level scheme illustrated in Fig.~\ref{fig:method_hierarchical},
where $T$ is restricted to be a multiple of 4:

\begin{gather}
W(R,T) = \langle \textbf{L}(0)_{\alpha\gamma} \{ \cdots
\textcolor{green}{\biggl[}
\textcolor{red}{\Bigl[}
\textcolor{blue}{[}
\textcolor{black}{\textbf{T}(R,n a)}
\textcolor{blue}{][}
\textcolor{black}{\textbf{T}(R,(n+1)a)}
\textcolor{blue}{]}
\textcolor{red}{\Bigr]} \notag \\
\textcolor{red}{\Bigl[}
\textcolor{blue}{[}
\textcolor{black}{\textbf{T}(R,(n+2)a)}
\textcolor{blue}{][}
\textcolor{black}{\textbf{T}(R,(n+3)a)}
\textcolor{blue}{]}
\textcolor{red}{\Bigr]}
\textcolor{green}{\biggr]}
\cdots  \}_{\alpha\beta\gamma\delta}
 \textbf{L}(T)^{*}_{\beta\delta} \rangle \label{eq:method_wilson_loop_multilevel}
\end{gather}

As a result of a rough optimisation process in the error reduction,
based on minimising the CPU time versus the error,
we choose the following parameters (see also \cite{Majumdar:2002}): The innermost averages
$\textcolor{blue}{[}
\textcolor{black}{\textbf{T}(R,n a)}
\textcolor{blue}{]}$ are calculated from 10 sets of time-like \emph{Multihit}-links,
each obtained after $n_1=10$ updates. The updates alternate heatbath and overrelaxation steps in the proportion 1:4.
The second-level averages,
$\textcolor{red}{\Bigl[}
\textcolor{blue}{[}
\textcolor{black}{\textbf{T}(R,n a)}
\textcolor{blue}{][}
\textcolor{black}{\textbf{T}(R,(n+1)a)}
\textcolor{blue}{]}
\textcolor{red}{\Bigr]}$
are calculated from $n_2=160$ averages of $\textcolor{blue}{[}
\textcolor{black}{\textbf{T}(R,na)}
\textcolor{blue}{]}$ and $\textcolor{blue}{[}
\textcolor{black}{\textbf{T}(R,(n+1)a)}
\textcolor{blue}{]}$, separated by
200 updates of the spatial links on time-slice $(n+1) a$.
Finally, the third-level averages are calculated from $n_3=165$
second-level averages separated by 200 updates of the spatial
links on time-slice $(n+2) a$.\\

$n_1=10$ seems rather small, but can be explained using
the confinement-deconfinement phase transition.
For the ratio $\frac{T_c}{\sqrt{\sigma}}$, \cite{Engels:1996dz} finds a value
\be
    \frac{T_c}{\sqrt{\sigma}}= 1.065(6)
\ee
with periodic boundary conditions. The corresponding critical temporal extent
(in lattice units) for our coupling
$\beta=6.0$ is then
\be
    N_c^{(p)} = \frac{a(\beta=6.0)}{T_c} \approx 3.76
\ee
As a rule of thumb, one can estimate that a slice with fixed boundary conditions with a
temporal extent of $\gtrsim \frac{1}{2} N_c^{(p)}$
will be "confined" \cite{Pepe}.
On the first level, we deal with
time slices of extent 1, a high temperature regime, corresponding to the deconfined phase.
Then the link-link-correlator has a large finite value, even at large $R$, and its determination is
easy: $n_1=10$ Multihit-averages are sufficient, and increasing $n_1$ further does
not reduce the final error as $1/\sqrt{n_1}$. On the next level, time slices
of extent 2
are in the confined phase. The signal then decreases
exponentially at large $R$.
We adjust $n_2$ so that the signal to noise ratio is about 1 for the distance $R=13 a$ which
we are interested in. On the third level,
the situation becomes more complicated and the best choice of all three parameters can only
be found using optimisation, to minimise CPU time versus error. We find that a three-level
scheme is more efficient than a one- or two-level scheme. \\

In Eq.(\ref{eq:method_wilson_loop_multilevel}), we have a product of $N$ tensors, the
three-level link-link-correlators with a temporal extent $4a$.
Just like in the case of the Multihit-method, the variance of each
tensor is reduced by a factor $\delta(R)$.
Thus, the variance of the Wilson loop average can be reduced
by as much as $\delta(R)^N$ for an effort $\propto N \delta(R)^{-2}$.
Variance reduction exponential in $N=T/4$ is achieved.

\subsubsection{Improved spatial transporter}
\label{subsec:method_improved_spatial_transporter}

Using the above technique, we are able to reduce exponentially the error coming from the temporal links of the Wilson loop.
But there is still the intrinsic noise, coming from the frozen spatial links at time-slices $T=0 \mod 4$,
which is relatively large \cite{Majumdar:2002}. How can we decrease it? This is what we do:
To provide additional error reduction also in the spatial transporter, we replace the spatial transporters
$\textbf{L}(0)_{\alpha\gamma}$ and
$\textbf{L}(T)^{*}_{\beta\delta}$ with staple-shaped transporters, which are
constructed in the following way (see Fig.~\ref{fig:method_improved_spatial}):
\begin{figure}[!t]
  \centering
  \psfrag{[[T(0)][T(a)]]}{\tiny
\textcolor{red}{\hspace{-0.1cm}$\Bigl[$}
\textcolor{blue}{\hspace{-0.075cm}$[$}
\textcolor{black}{\hspace{-0.075cm}$\textbf{T}(0)$}
\textcolor{blue}{\hspace{-0.075cm}$][$}
\textcolor{black}{\hspace{-0.075cm}$\textbf{T}(a)$}
\textcolor{blue}{\hspace{-0.075cm}$]$}
\textcolor{red}{\hspace{-0.1cm}$\Bigr]$}
}
\psfrag{[[T(2a)][T(3a)]]}{\tiny
\textcolor{red}{\hspace{-0.1cm}$\Bigl[$}
\textcolor{blue}{\hspace{-0.075cm}$[$}
\textcolor{black}{\hspace{-0.075cm}$\textbf{T}(2a)$}
\textcolor{blue}{\hspace{-0.075cm}$][$}
\textcolor{black}{\hspace{-0.075cm}$\textbf{T}(3a)$}
\textcolor{blue}{\hspace{-0.075cm}$]$}
\textcolor{red}{\hspace{-0.1cm}$\Bigr]$}
}
\psfrag{adjointspatialtransporter}{ \tiny
\textcolor{magenta}{adjoint spatial transporter}
}
  \mbox{\includegraphics[width=12.0cm]{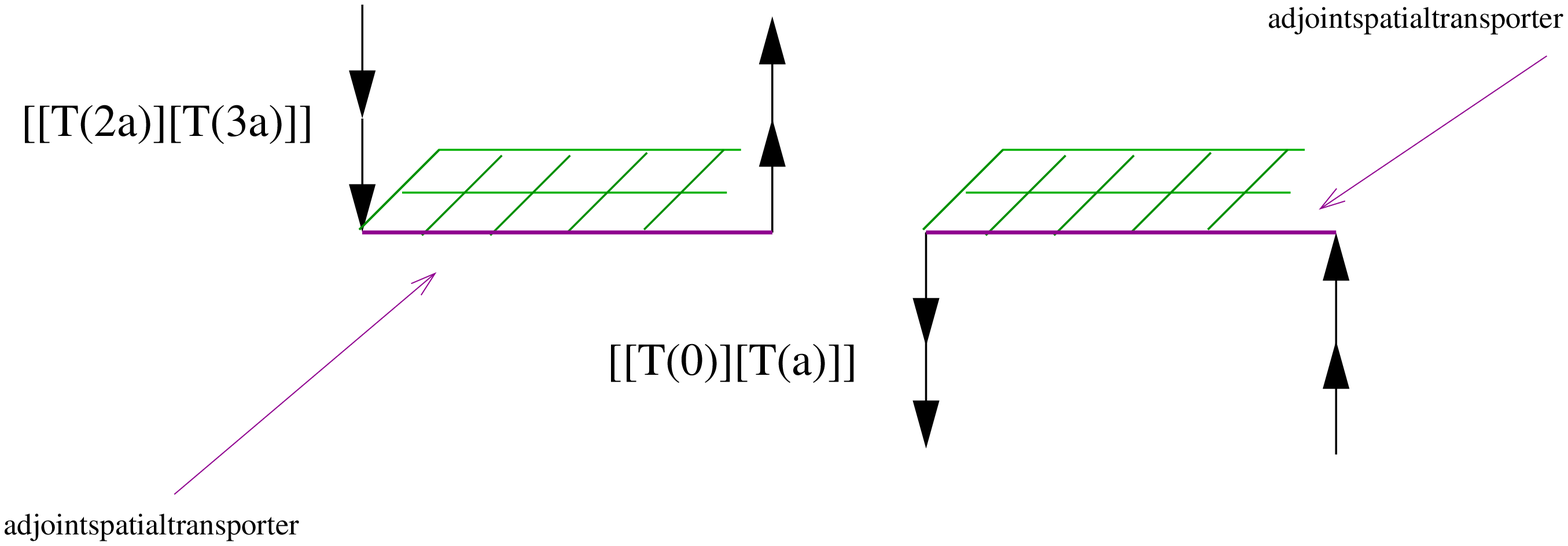}}
  \caption{The improved spatial transporters. After each calculation of
    second-level averages $\textcolor{red}{\Bigl[}
\textcolor{blue}{[}
\textcolor{black}{\textbf{T}(R,n a)}
\textcolor{blue}{][}
\textcolor{black}{\textbf{T}(R,(n+1)a)}
\textcolor{blue}{]}
\textcolor{red}{\Bigr]}$, we form staple-shaped transporter
    including smeared spatial links at the time-slice $(n+2)a$.}
  \label{fig:method_improved_spatial}
\end{figure}

$(i)$ After each calculation of
second-level averages, denoted as \includegraphics[height=0.5cm]{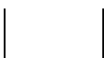}\;, we form
the smeared spatial links,
\includegraphics[height=0.15cm]{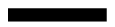}$\;$, at
time-slice $(n+2)a$. $(ii)$ We multiply them with the second-level
averages to obtain the staple-shaped transporter
\includegraphics[height=0.5cm]{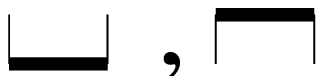}.
$(iii)$ This procedure is repeated $n_3=165$ times, each time after updating the spatial
links on time-slice $[(n+2) a]$ during the calculation of the third-level average.
These error-reduced staple-shaped transporters replace the naive spatial transporters
$\textbf{L}(0)_{\alpha\gamma}$ and $\textbf{L}(T)^*_{\beta\delta}$ and by contracting them with
none, one, two, etc... third-level-averages one obtains Wilson loops at a fixed $R$ with
temporal extent $T=4,8,12,$ etc...

\subsection{Diagonalisation procedure}
\label{subsec:method_diagonalisation_procedure}

Given a set of states $\phi_i$ and the correlation matrix $V_{ij}(R,T)$
as introduced in Section \ref{sec:staticpotential}, one can approximate
the eigenstates correlation matrix $\Gamma$ defined in Eq.(\ref{eq:staticpotential_gamma_matrix}),
get information on the eigenstates $\Psi^{(n)}$ and
extract the lowest energies $E_n$ using a diagonalisation procedure.\\

For a given separation $R$, the correlation matrix is defined in Eq.(\ref{eq:staticpotential_correlation_matrix}) as
\be
    V_{ij}(R,T)  =  \langle \phi_j(R) \mid {\bf \hat{T}}^T \mid \phi_i(R) \rangle \;.
\ee

A naive determination of the lowest energies $E_n$ is obtained by looking for a plateau in the ratio of
eigenvalues $\frac{\lambda^{(n)}(R,T)}{\lambda^{(n)}(R,T+1)}$ of $V_{ij}(R,T)$
for increasing $T$ (see Eq.(\ref{eq:staticpotential_diagonalisation_eigenvalues})):
\be
    E_n(R,T) = \lim_{T \to \infty} \ln \frac{ \lambda^{(n)}(R,T) }{ \lambda^{(n)}(R,T+1) } \;.
\ee
This simple method works very well, especially in the multichannel Ansatz.
Nevertheless, we want to increase the signal of the desired state as much as possible.
For a finite basis, for small $T$, the eigenstates change with $T$. To improve $T$-convergence,
we apply variational diagonalisation which consists of solving the generalised eigenvalue problem
\be \label{eq:method_diagonalisation}
    V_{ij}(R,T) \; v^{(n)}_j(R,T,T_0) = \lambda^{(n)}(R,T,T_0) \; V_{ij}(R,T_0) \; v^{(n)}_j(R,T,T_0)\;,\;T>T_0
\ee
The eigenvalues $\lambda^{(n)}(R,T,T_0)$ also fulfill Eq.(\ref{eq:staticpotential_diagonalisation_eigenvalues}) but their
coefficients $f^{(n)}(R)$ are enhanced by construction, compared
to the previous ones.
Once the eigenvectors ${\bf v}^{(n)}$ and eigenvalues $\lambda^{(n)}$ are known, one may
approximate the eigenstates as a superposition
of the operator states using
\be \label{eq:method_normalisation}
    \Psi^{(n)}(R,T) = N^{(n)}(R,T,T_0) \sum_j v^{(n)}_j(R,T,T_0) \phi_j(R,T) \equiv \sum_j a^{(n)}_j(R,T,T_0) \phi_j(R,T)
\ee
where the constants $N^{(n)}$ are chosen such that the $\Psi^{(n)}$ are normalised to unity.\\

It is well known from the literature, that the groundstate energy can be extracted nicely
using a single-mass Ansatz even in the broken-string regime $R>R_b$,
starting with $T_0=0$, \emph{if} broken-string state operators are included in the multichannel (variational) basis.
Indeed, we can confirm this observation.
The situation is different when one uses a pure Wilson loop operator basis, at $R>R_b$.
Because the overlap of the Wilson loop with the broken-string state is so weak, one must
choose $T_0$ very large to ensure that the lowest-lying eigenstates at $T_0$ and $T$ are
both broken-string states. Then, the high sensitivity of Eq.(\ref{eq:method_diagonalisation}) to
statistical noise renders the analysis delicate: The matrix $V_{ij}(R,T_0)$ may not be invertible.
As a trade-off, we choose a small value of $T_0$, $T_0=4 a$, but must use a two-mass Ansatz to
account for all data points.

\subsection{Gluelumps}
\label{subsec:method_gluelumps}
\begin{figure}[!t]
  \centering
  \psfrag{C1}{${\bf C}_1$}
  \psfrag{C2}{${\bf C}_2$}
  \psfrag{C3}{${\bf C}_3$}
  \psfrag{C4}{${\bf C}_4$}
  \mbox{\includegraphics[width=12.0cm]{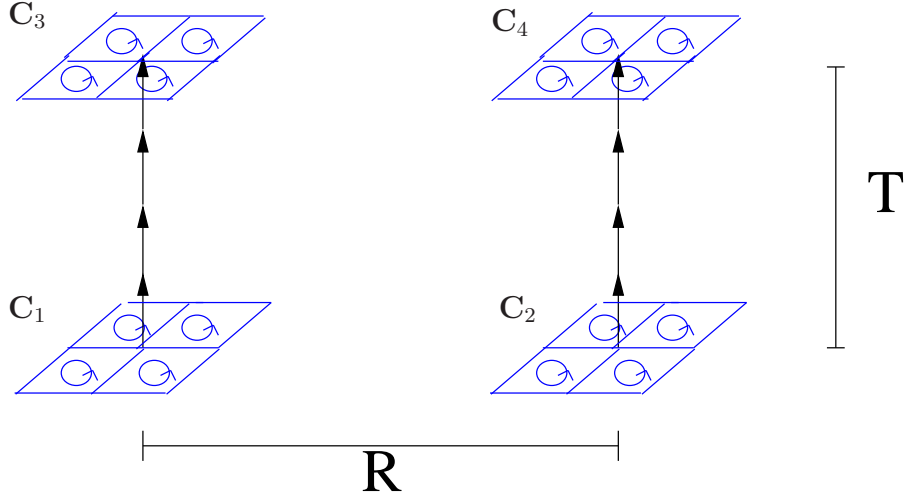}}
  \caption{Interacting gluelumps at distance $R$. Four "clovers" are stuck on a
  link-link-correlator tensor of temporal extent $T$.}
  \label{fig:method_gluelump_inter_c}
\end{figure}
In order to fully implement the multichannel Ansatz, we must
consider the two-gluelump correlator, which is in the notation of Eq.(\ref{eq:staticpotential_multichannel_ansatz})
\begin{equation}
    \vcenter{\vspace{-2ex}\hbox{\includegraphics{pictures/GG_small.eps}}} =  GG(R,T) \;.
 \end{equation}
The broken-string state can be described by the presence of two gluelumps, each formed
by the coupling of adjoint glue to an adjoint static charge \cite{Michael:1985}.
It is of interest to measure the mass and the correlator of gluelumps. A simple way to probe the gluon field distribution,
denoted as \includegraphics{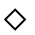},
around the adjoint static charge is the so-called clover-discretisation of $F_{\mu\nu}$,
denoted as ${\bf C}_{m,\mu\nu}={\bf C}_{1,\mu\nu},...,{\bf C}_{4
,\mu\nu}$ in Fig.~\ref{fig:method_gluelump_inter_c}.
\begin{align}
    {\bf C}_{m,\mu\nu} &=\frac{1}{4} \sum_{i=1}^4 \textnormal{Smeared Spatial Plaquette$(m,\mu,\nu)_i$} \notag \\
       &= C_{m,\mu\nu}^0 \mathbbm1_{2\textnormal{x}2} + i C_{m,\mu\nu}^\alpha \sigma^\alpha \label{eq:method_clover_observable}
\end{align}
The anti-hermitian part of the clover ${\bf C}_a $ approximates $F_{\mu\nu}$
\begin{equation}
    \frac{1}{2 i}({\bf C}_{m,\mu\nu} - {\bf C}_{m,\mu\nu}^\dagger)  =   C_{m,\mu\nu}^\alpha \sigma^\alpha   = g_0 a^2 F_{\mu\nu} + O(a^4)
\end{equation}
where the $\sigma^\alpha$ are the Pauli matrices and we average over
 the four oriented spatial plaquettes
that share a corner with one end of the time-like line. We choose the orientation of
Fig.~\ref{fig:method_gluelump_inter_c} (all four staples clockwise)
to obtain the lowest-lying gluelump
mass \cite{Poulis:1995nn}. The plaquettes are built using
fundamentally smeared links (as described in Eq.(\ref{eq:method_smearing_fundamental})). \\

To measure the gluelump mass, one considers only one gluelump,
e.g. the left side
(${\bf C}_1$ - ${\bf C}_3$) in Fig.~\ref{fig:method_gluelump_inter_c}.
The gauge-invariant operator $CC(T)$ is an adjoint time-like line of length $T$
using Eq.(\ref{eq:method_adjointlink}), $A^{\alpha\beta}(T)$,
which is coupled at the two ends to the clovers (${\bf C}_1$ and ${\bf C}_3$).
\be
    CC(T) = C_1^\alpha(0) A^{\alpha\beta}(T) C_3^\beta(T)
\ee

The adjoint time-like line can be measured using the Multihit-method or by applying the
Multilevel-idea to this particular problem.
The mass $M(Qg)$ can then be extracted using
\be
    CC(T) \sim e^{-M(Qg) T}
\ee
The overlap with the lowest-lying state is enhanced by using
smeared links to build the clover observable Eq.(\ref{eq:method_clover_observable}).
We would like to mention at this point
that the gluelump mass by itself has no real physical meaning,
since it contains a UV-divergence
in the continuum limit due to the self-energy of the time-like links.
Only the difference between this divergent mass and
another similarly divergent one, like the static potential, makes physical sense.
As a consequence, the string breaking distance
$R_b$ remains constant in physical units,
while the energy of the level-crossing diverges as $\beta \rightarrow \infty$. \\

To measure the correlation of two gluelumps, one has to consider the full object in
Fig.~\ref{fig:method_gluelump_inter_c}. The four clovers ${\bf C}_1,...,{\bf C}_4$ are
 measured as described above. The correlation $GG(R,T)$ of two gluelumps separated by a
distance $R$ can be measured by contracting the four clovers to the link-link-correlator tensors
with temporal extent $T$, $\textbf{T}(R,T)$. The same tensors, obtained with the Multilevel algorithm
and used for Wilson loops, are also used here.
\be
    GG(R,T) = C_2^\gamma(0) C_1^\alpha(0)
        \textbf{T}_{\alpha\beta\gamma\delta}(R,T) C_3^\beta(T) C_4^\delta(T)
\ee
$GG(R,T)$ can be used in two ways: On the one hand, as mentioned in the beginning of this Section,
it is a contribution to the multichannel matrix; on the other hand, we can try to extract, from it alone,
the energies of the unbroken-string and of the broken-string states since presumably the
two-gluelump correlator has projection on both states.
\be \label{eq:method_gluelump_double_exp}
     GG(R,T) \sim g_0 e^{-V(R) T} + g_1 e^{-E_1(R) T}\;,\;T > T_{\textnormal{min}}
\ee
where $V(R)$ is the static potential and $E_1(R)$ the first excited state energy.
The operator has obviously a good overlap with the broken-string state,
but a poor one with the unbroken-string state.
This situation mirrors that of the Wilson loop, described
in the beginning of Section \ref{sec:approaches}.
For $R<R_b$, the unbroken-string state
is the groundstate. Therefore $g_0$ is expected to be
small compared to $g_1$, and the first excited state, with the energy of two gluelumps,
is dominating for small $T$. Since the groundstate
will be visible for large temporal extents only, we need a two-mass Ansatz to describe the correlator.
At large distances, $R \geq R_b$, the broken-string state is the groundstate and
also the dominating one ($g_0 \gg g_1$),
therefore a single-mass $g_0 e^{-V(R) T}$ will suffice. \\

In the case of Wilson loops, we attach improved spatial transporter to the link-link-correlators.
Here, we use non-improved clovers for simplicity. Therefore, we have more statistical noise, which makes it difficult to
extract the groundstate, if the turning point is large. This is the case, for distances $R$ close to but
below the string breaking distances $R_b$. For details see Subsection \ref{subsec:results_gluelumps}.

\subsection{Multichannel Ansatz}
\label{method_multichannel_ansatz}
To complete the multichannel Ansatz, we must also consider the off-diagonal
elements in Eq.(\ref{eq:staticpotential_multichannel_ansatz}), denoted $S_{s_1} G(R,T)$ and $G S_{s_2}(R,T)$.
\begin{align}
    S_{s_1} G(R,T) &= \textbf{L}_{s_1}(0)_{\alpha\gamma}
        \textbf{T}_{\alpha\beta\gamma\delta}(R,T) C_3^\beta(T) C_4^\delta(T) \\
    G S_{s_2}(R,T) &= C_2^\gamma(0) C_1^\alpha(0)
        \textbf{T}_{\alpha\beta\gamma\delta}(R,T) \textbf{L}_{s_2}(T)^*_{\beta\delta}
\end{align}

To extract their values at $T=0 \mod 4$, we use the non-improved spatial transporter $\textbf{L}_{s_1}(0)$
and $\textbf{L}_{s_2}(T)$, where we smeared the links beforehand using $s_1$ respectively $s_2$
smearing iterations. The clovers are denoted as ${\bf C}_1,...,{\bf C}_4$. We attach them
to the link-link-correlators $\textbf{T}(R,T)$.
The complete multichannel matrix is
\begin{align}
    V_{ij}(R,T)  &=  \left(
    \begin{matrix}
        \includegraphics{pictures/WW_small.eps} & \includegraphics{pictures/WG_small.eps} \\
        \includegraphics{pictures/GW_small.eps} & \includegraphics{pictures/GG_small.eps}
    \end{matrix}
    \right)
    =
    \left(
    \begin{matrix}
        S_{s_1} S_{s_2} (R,T) & S_{s_1} G(R,T) \\
        G S_{s_2}(R,T) & GG(R,T)
    \end{matrix}
    \right) \notag \\
    & =
    \left(
    \begin{matrix}
        S_{15} S_{ 15 }& S_{15} S_{ 30 }& S_{15} S_{ 60 }& S_{15} G\\
        S_{30} S_{ 15 }& S_{30} S_{ 30 }& S_{30} S_{ 60 }& S_{30} G\\
        S_{60} S_{ 15 }& S_{60} S_{ 30 }& S_{60} S_{ 60 }& S_{60}G\\
        G  S_{15}  & G  S_{30}  & G  S_{60}  & G  G
    \end{matrix}
    \right) \label{eq:method_multichannel_matrix}
 \end{align}

This $4 \times 4$ matrix can be analysed using the diagonalisation procedure
 described in  Subsection \ref{subsec:method_diagonalisation_procedure}. We end up
 with enhanced signals for the three lowest-lying states, plus effective information
 about higher states.

\subsection{Polyakov loops}

 According to
Eq.(\ref{eq:approaches_polyakov}) we can
extract a temperature-dependent potential $V_{\cal{T}}(R)$. The correlator of
two adjoint Polyakov loops can
 be easily built by using the link-link correlator tensors and the
 tensor-multiplication defined
in Eq.(\ref{eq:method_tensor_multiplication}).
\begin{align} \label{eq:method_polyakov_tensor}
    \langle P(0) P^*(R) \rangle = \langle\{ \cdots
\textcolor{green}{\biggl[} &
\textcolor{red}{\Bigl[}
\textcolor{blue}{[}
\textcolor{black}{\textbf{T}(R,n a)}
\textcolor{blue}{][}
\textcolor{black}{\textbf{T}(R,(n+1)a)}
\textcolor{blue}{]}
\textcolor{red}{\Bigr]}\notag \\
&
\textcolor{red}{\Bigl[}
\textcolor{blue}{[}
\textcolor{black}{\textbf{T}(R,(n+2)a)}
\textcolor{blue}{][}
\textcolor{black}{\textbf{T}(R,(n+3)a)}
\textcolor{blue}{]}
\textcolor{red}{\Bigr]}
\textcolor{green}{\biggr]}
\cdots  \}_{\alpha\alpha\gamma\gamma} \rangle
 \notag \\
        &= \; e^{-V_{\cal{T}}(R)/\cal{T}}
\end{align}

\section{Results}
\label{sec:results}

We are using a $3d$-lattice with extent $(48 a)^2 \times 64 a$ at inverse coupling
$\beta=\frac{4}{a g^2}=6.0$. The lattice spacing $a$ can be obtained
from the Sommer scale $r_0$\footnote{The phenomenological interpretation of this scale is
valid only for QCD. When extracting the lattice spacing $a$ from the Sommer scale $r_0=0.5$ fm,
the resulting value of $a$ depends on the Ansatz chosen for the potential, and on the
fitting range for the force. We chose Ansatz Eq.(\ref{eq:results_VR_Ansatz}),
which includes a perturbative logarithmic term, because not including this term causes an
unacceptably bad fit.
Fitting the force over the interval $3 a \leq R \leq 7 a$, one
obtains $r_0/a$ is $4.890(1)$, and $a=0.1022(1)$ fm. This value
changes by $0.5 \%$ under a variation of the fitting interval.
Similar ambiguities arise when one tries to extract the lattice
spacing from the string tension ($\sigma a^2$). Setting
$\sqrt{\sigma}=440 \textnormal{MeV} \approx (0.45 \textnormal{fm})^{-1}$,
one obtains $a=0.1136(1)$ fm, with a
systematic variation of about $0.5 \%$ with the fitting range.}
\cite{Sommer:1993ce}, which is defined by
\be
    r_0^2 F_{\textnormal{fund}}(r_0)=1.65\;.
\ee
Setting $r_0 = 0.5$ fm and comparing with the lattice result for $r_0/a$, one obtains $a = 0.1022(1)$ fm.
A description of our procedure to extract the fundamental force $F_{\textnormal{fund}}$
is given in Subsection \ref{subsec:results_casimir}.\\

We present our results in the following order:
\begin{enumerate}
\item  The static potential and excited states extracted from
Wilson loops only.
\item The static potential extracted from the two-gluelump correlator.
\item The static potential and excited states obtained from the multichannel Ansatz.
\item The temperature-dependent potential obtained from the Polyakov loop correlators.
\item The comparison of fundamental and adjoint potentials and the issue of Casimir scaling.
\end{enumerate}
We have analysed 44 configurations, which appear to be statistically uncorrelated.
To extract the statistical errors we apply the jackknife method.

\subsection{Wilson loops only}
\begin{figure}[!t]
  \centering
  \mbox{\includegraphics[angle=-90,width=12.0cm]{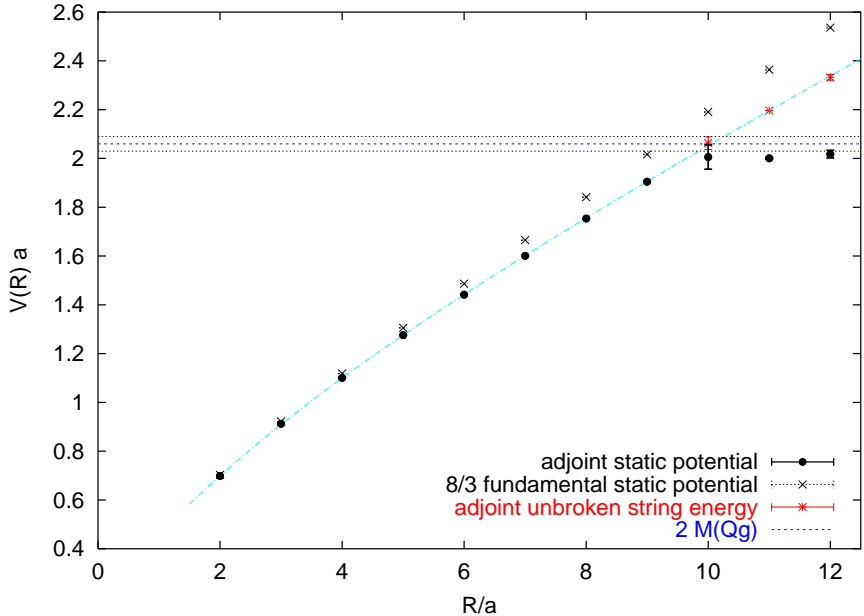}}
  \caption{The adjoint and fundamental static potentials $V(R)$ (the latter multiplied by the Casimir
  factor $\frac{8}{3}$) versus $R$ using Wilson loops only.
The adjoint static potential
remains approximately constant for $R \geq R_b \approx 10 a$ proving string breaking.
The unbroken-string state energy is also drawn. The horizontal line at $2.06(3) a^{-1}$ represents
twice the mass of a gluelump. }
  \label{fig:results_stringbreaking_with_fund}
\end{figure}
We measure both the fundamental and the adjoint potential between two static
charges.
In the first case, string breaking cannot occur since the system does not contain particles
which can screen charges in the fundamental representation. Nevertheless, we can compare
our results with accurate data available in the
literature \cite{Born:1994,Poulis:1995nn}. We will also need these values later on, to discuss
the issue of Casimir scaling (see Subsection \ref{subsec:results_casimir}).
In the case of adjoint static charges, string breaking should occur.
A summary of our results
is given in Fig.~ \ref{fig:results_stringbreaking_with_fund},
where we show
the fundamental potential multiplied by the Casimir ratio $\frac{8}{3}$
(see Eq.(\ref{eq:method_pert_staticpotential}))
and the adjoint static potential. It can clearly be seen that the adjoint
static potential
remains approximately constant for $R \geq 10 a$ proving string breaking at $R_b\approx 10 a$.
The unbroken-string potential is also shown.

The horizontal line at $2.06(3) a^{-1}$ represents
twice the mass of a gluelump, whose evaluation is described in
Subsection \ref{subsec:method_gluelumps}.
This is the expected value of the static potential of the system,
when the string is broken, since the broken-string state is modelled
by the presence of two gluelumps whose interaction is screened.

Excited states are not visible for the fundamental case
since the shortest Wilson loops we consider have
a minimal temporal extent of $T=4 a$ and excited states are already strongly suppressed.
But they are clearly seen in the adjoint case for distances larger than the
string breaking distance since the Wilson loop has very good overlap
with the unbroken-string state which is an excited state for $R > R_b$. More about excited
states can be found in Subsection \ref{subsec:results_excited_states}.

\subsubsection{Static potential}
\label{subsec:results_staticpotential}

We start our discussion with the extraction of the fundamental static potential $V_{\textnormal{fund}}(R)$.
We consider only one level of fundamental smearing (30 iterations of Eq.(\ref{eq:method_smearing_fundamental}))
of the fundamental spatial links and do not consider a Wilson loop matrix in the sense of Eq.(\ref{eq:method_wilson_loop_matrix}).
A single-mass Ansatz works nicely at all $R$ in the temporal
range $T_{\textnormal{min}}=12a \leq T \leq 60a$, where we have no measurable contribution of excited states.
The extracted $V_{\textnormal{fund}}(R)$ is in full agreement with the literature.

From the static potential we can extract the string tension $\sigma$. This gives us a crosscheck with previous
determinations \cite{Teper:1998te} and a way to express the lattice spacing in physical units.
We use a string-motivated Ansatz
\be \label{eq:results_VR_Ansatz}
V(R) \sim V_0 + l \ln \frac{R}{a} -\frac{\gamma}{R}+ \sigma R
\ee
The Coulombic $\ln \frac{R}{a}$ term follows from $3d$ perturbation theory (see Eq.(\ref{eq:method_pert_staticpotential})).
The $1/R$ term follows from the bosonic string model. $\gamma$ is a universal constant
with value $\gamma=\frac{\pi}{24}(d-2)$ in $d$ dimensions \cite{Luscher:1981}.
The linear term describes confinement, and $\sigma$ is the string tension.\\
We fit all parameters and find for the string tension $\sigma = 0.0625(5) a^{-2}$. This value is stable
and in full agreement with \cite{Teper:1998te}. Using our Ansatz Eq.(\ref{eq:results_VR_Ansatz}),
$\gamma$ cannot be reliably extracted by a global fit of the static potential. Using instead\footnote{We do observe an increase in $\gamma$
with increasing $R$, visible until $R \sim 6a$.
This increase can be understood as a
 $1/R$-correction to $\gamma$ coming from the next-to-leading
term in the bosonic string theory \cite{Luscher:1981}.}
\be
 \gamma = -  \frac{\partial^2 V(R)}{\partial R^2} R^3  \;,
\ee
the extracted $\gamma$ tends to the universal value $\frac{\pi}{24} \approx 0.131$: $\gamma \underset{R=6a}{=} 0.126(12)$ and remains
stable for $R \geq 6 a$ albeit with larger errors.\\
In the following case of the adjoint static potential, the Ansatz Eq.(\ref{eq:results_VR_Ansatz}) does
not result in stable parameters, with or without the Coulombic $\ln \frac{R}{a}$ term. Nevertheless, we include
in Fig.~\ref{fig:results_stringbreaking_with_fund}
a best fit of the adjoint unbroken-string energy in the range $2 a\leq R \leq R_b$ using this Ansatz.\\

\begin{figure}[!t]
  \centering
  \psfrag{DATAPOINTATBETASIX}{\LARGE \textcolor{red}{data points at $\beta=6.0$} \normalsize}
  \mbox{\includegraphics[angle=-90,width=12.0cm]{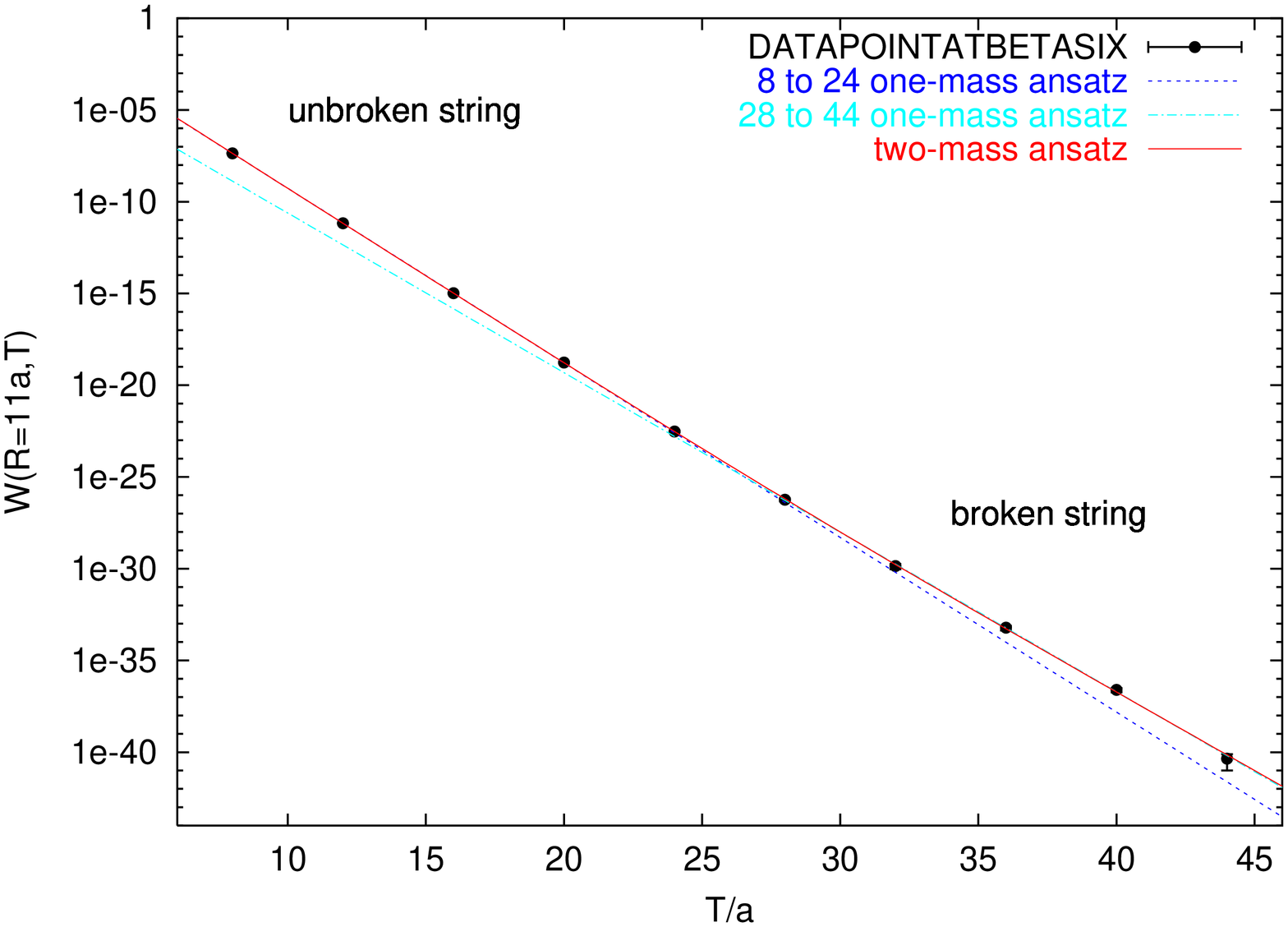}}
  \mbox{\includegraphics[angle=-90,width=12.0cm]{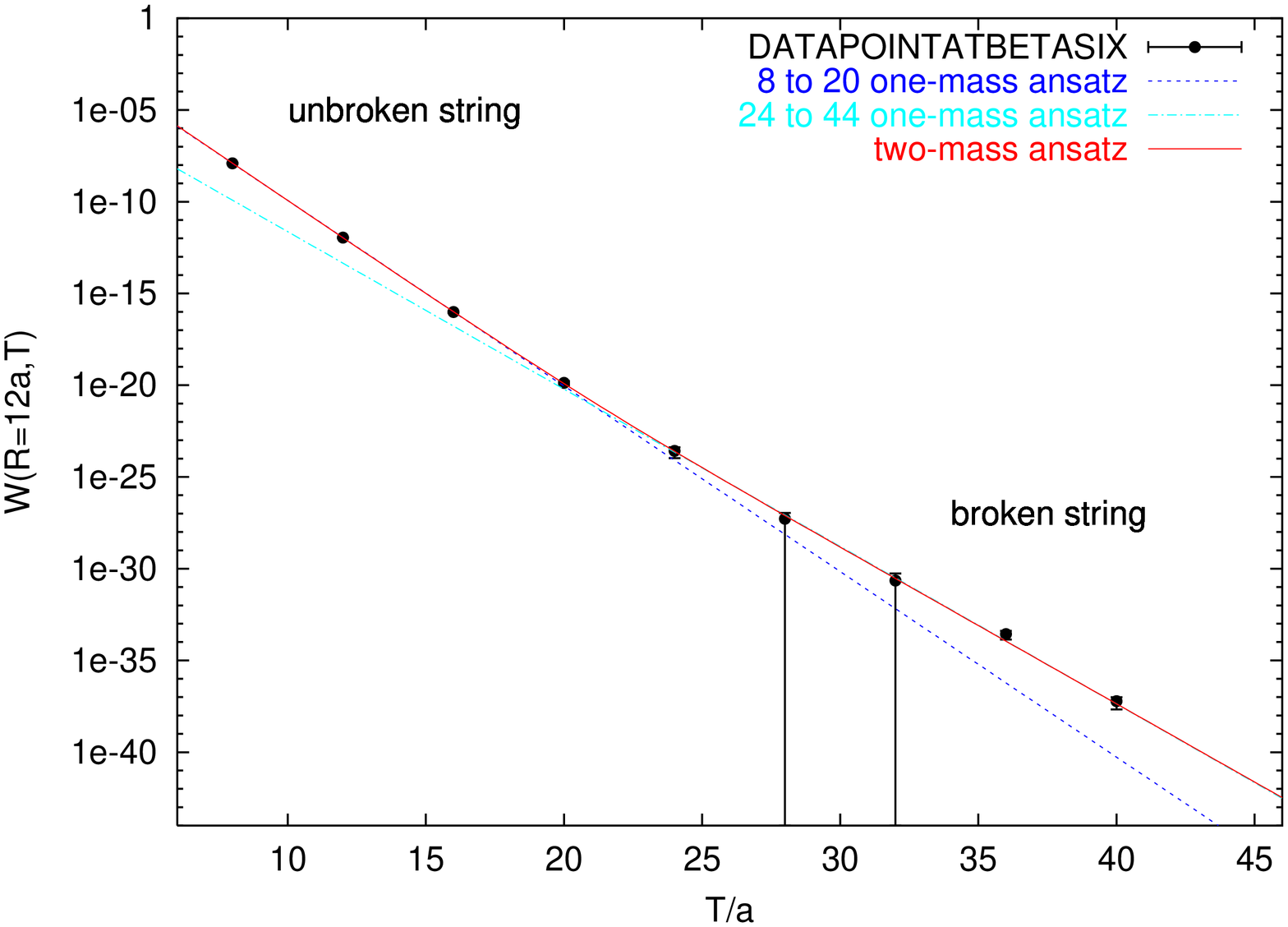}}
  \caption{Adjoint Wilson loop data versus $T$, for $R=11 a$ and $R=12 a$, obtained from a diagonalisation
  procedure Eq.(\ref{eq:results_simplified_version}), applied to Wilson loops, considering three different levels of smearing.
  A two-mass Ansatz accounts for all data points. Single-exponentials (dotted lines) do not.
  At large $T$, the broken-string groundstate is exposed.
  Note how small a signal can be measured.}
  \label{fig:results_two_mass_ansatz}
\end{figure}
The extraction of the adjoint static potential works well using a
single-mass Ansatz for $T \geq 4$ and $R < R_b$. At larger distances $R$, it is a
more complicated matter since the string breaks and the Wilson loop
has a poor overlap with the broken-string. This makes the two-mass Ansatz mandatory.
To extract the energies of the groundstate and first excited state, as shown in the figures,
we use the diagonalisation procedure described in
Subsection \ref{subsec:method_diagonalisation_procedure}.
To illustrate that the static potential $V(R)$ at a fixed $R > R_b$ cannot been determined
by a single-mass, we show in Fig.~\ref{fig:results_two_mass_ansatz}
$W(R,T) = \lambda^{(0)}(R,T,T_0)$ at $R=11 a$ and $R=12 a$.
Here we want to make use of the full Wilson loop data without distorting the ratio
 $\frac{c_1}{c_0}$ of Eq.(\ref{eq:staticpotential_two_mass_ansatz}).
We use a simplified version of the diagonalisation procedure to obtain $\lambda^{(0)}(R,T,T_0)$
using Wilson loops only:

We have three types of staple-shaped transporter \includegraphics[height=0.25cm]{pictures/twolevelhats.eps}
as used in Eq.(\ref{eq:method_multichannel_matrix}). In the same notation, the Wilson loops correlation
matrix is
\begin{equation} \label{eq:method_wilson_loop_matrix}
    \vcenter{\vspace{-2ex}\hbox{\includegraphics{pictures/WW_small.eps}}} =  S_{s_1} S_{s_2}(R,T)  =  \left(
    \begin{matrix}
        S_{15} S_{ 15 }& S_{15} S_{ 30 }& S_{15} S_{ 60 }\\
        S_{30} S_{ 15 }& S_{30} S_{ 30 }& S_{30} S_{ 60 }\\
        S_{60} S_{ 15 }& S_{60} S_{ 30 }& S_{60} S_{ 60 }
    \end{matrix}
    \right)
 \end{equation}

\begin{enumerate}
\item For a fixed $R$, we diagonalise the matrix $S_{s_1} S_{s_2}(R,T_0)$, where we choose $T_0$ so that
the overlap with the desired state (e.g. the groundstate) is as large as possible and the
signal still quite accurate. Setting $T_0 \gtrsim T_P$ is a natural choice. E.g. at $R=12 a$ we choose $T_0=24 a$.
\item We use the eigenvectors ${\bf v}_0(R,T_0)$,
${\bf v}_1(R,T_0)$ and ${\bf v}_2(R,T_0)$, where the corresponding eigenvalues fulfill
$\lambda^{(0)}(R,T_0) > \lambda^{(1)}(R,T_0) >  \lambda^{(2)}(R,T_0)$, to project $ S_{s_1} S_{s_2}(R,T) $ to the
different states at all $T$ by
\be \label{eq:results_simplified_version}
    \lambda^{(n)}(R,T,T_0) = v_{n,{s_1}}(R,T_0) \; S_{s_1} S_{s_2}(R,T) \; v_{n,{s_2}}(R,T_0)
\ee
\end{enumerate}

$\lambda^{(0)}(R,T,T_0)$, the largest eigenvalue, contains amplified information
about the groundstate and
$\lambda^{(1)}(R,T,T_0)$ information about the
first excited state.
$\lambda^{(2)}(R,T,T_0)$  is some effective value containing information
about the remaining excitations.
$W(R,T) = \lambda^{(0)}(R,T,T_0)$ has to be analysed with a two-mass Ansatz and
the ratio $\frac{c_1}{c_0}$ can be extracted.
\be
   W(R,T) =  c_0 e^{-V(R) T} + c_1 e^{- E_1 T}\;,\;T > T_{\textnormal{min}}\;.
\ee

$V(R)$ corresponds to the groundstate, $E_1$ is the first excited state energy.
The groundstate is the broken-string state and therefore, $V(R)$ should be
the energy of two gluelumps $E(2 Q\bar{g}) \approx 2 M( Q\bar{g})$.
At small temporal extent $T$ of the Wilson loop $W(R,T)$ we get a
larger slope than at large $T$, as visible in Fig.~\ref{fig:results_two_mass_ansatz}.
This can be explained as elaborated in Section \ref{sec:approaches}:
At small $T$ the signal is dominated by the unbroken-string state. The broken-string state can only be
observed once $T$ is large enough since the Wilson loop observable has a poor overlap
with this groundstate.
The ratio $\frac{c_1}{c_0}$ quantifies the domination of the unbroken-string state signal versus
the broken-string state and is related to the turning point $T_P$ (see Eq.(\ref{eq:approaches_turningpoint})).
We obtain $\frac{c_1}{c_0}\underset{R = 12a}{=} 3.1(4) 10^3$, while the prediction of
the strong coupling expansion \cite{Drummond:1998ar} was $\frac{c_1}{c_0}\underset{R = 12a}{\approx}2\times 10^5$
and (see Eq.(\ref{eq:approaches_turningpoint_estimate})) $T_P^{(est)} \underset{R = 12a}{\approx} 42a$.
Our numerical determination of the turning point $T_P \underset{R = 12a}{=} 22(1)a$ is clearly below this estimate.

Note that we can detect signals down to $10^{-40}$, which corresponds to $10^{80}$
ordinary measurements. Previously, only signals
down to $10^{-7}$ have been measured, i.e. in a regime where the
unbroken-string state is dominating over the groundstate. This explains why
string breaking has not been observed in Wilson loops up to now.

\subsubsection{Excited states}
\label{subsec:results_excited_states}
\begin{figure}[!t]
  \centering
  \mbox{\includegraphics[angle=-90,width=12.0cm]{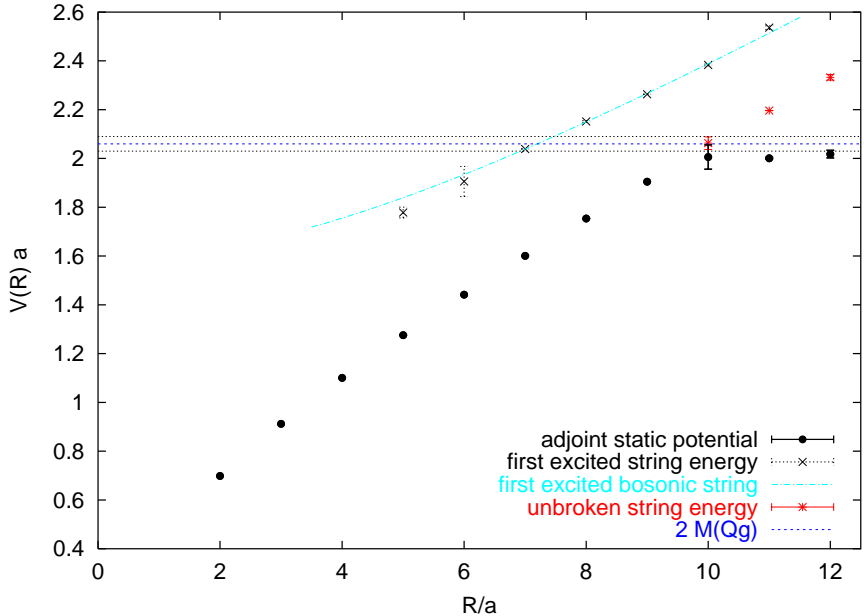}}
  \caption{The static adjoint potential $V(R)$ versus $R$
  (same as in Fig.~\ref{fig:results_stringbreaking_with_fund}) and the first excited
  unbroken-string state energy using Wilson loops only. We also show the energy $\tilde{E}_1(R)$
  (Eq.(\ref{eq:results_bosonic_string})) resulting from the
  relativistic Nambu string theory.
  The horizontal line at $2.06(3) a^{-1}$ represents twice the mass of a gluelump.}
  \label{fig:results_first_excited_state}
\end{figure}
Excited states of the fundamental representation are suppressed too much for us to measure.
But in the adjoint representation, we have clear information about the
first excited state. Using the diagonalisation procedure Eq.(\ref{eq:method_diagonalisation}),
one source of information is the two-mass fit
of $W(R,T) = \lambda^{(0)}(R,T,T_0)$ at large distances $R \geq R_b$ where the first
excited state is the unbroken-string state. Another, related, source of information,
also for smaller $R$, is $W_1(R,T)=\lambda^{(1)}(R,T,T_0)$.\\

In the same manner as for the static potential, we adopt here a two-mass Ansatz
\be
   W_1(R,T) =  e_0 e^{-V(R) T} + e_1 e^{-V_1(R) T}\;,\;T > T_{\textnormal{min}}\;.
\ee
where $V(R)$ is the static potential and $V_1(R)$ the energy of the first excited
unbroken-string state (as it turns out).
In Fig.~\ref{fig:results_first_excited_state},
for distances smaller than the string breaking
distance $R_b$ we see a clear signal of the
first excited state. However, at larger distance we expect contributions from at least
three states: The two-gluelump groundstate, the unbroken-string state, and an
excited unbroken- or broken-string state, since one expects
a level-crossing of the latter two depending on the spatial distance $R$.

Surprisingly, at $R=8 a$  and $R=9 a$ we completely miss the broken-string state.
The explanation lies in the spectrum at $R=8 a,9 a$. The first three terms entering the
expansion of the Wilson loop Eq.(\ref{eq:staticpotential_expansionwilsonloop}) are:
\begin{itemize}
\item $c_0 e^{-E_0 T }$ groundstate: lowest-lying energy state of the unbroken-string,
\item $c_1 e^{-E_1 T }$ first excited state: lowest-lying energy state of the broken-string,
\item $c_2 e^{-E_2 T }$ second excited state: first excited state of the unbroken-string.
\end{itemize}
The overlap of the Wilson loop operator with the second excited state (unbroken-string), $c_2$,  is larger than the overlap
with the first excited state, $c_1$. In addition, the difference between the two corresponding energies
$E_1$ and $E_2$ is small. In our case, the second excited state indeed dominates over the first in the accessible
range of Euclidean times. Therefore,
we miss the broken-string state at $R=8a$ and $R=9a$. \\

From our fit of the adjoint potential via Eq.(\ref{eq:results_VR_Ansatz}) we have extracted the
string tension $\sigma$. We then consider the relativistic Nambu string \cite{Perantonis:1989},
which predicts for the first excited string energy
\be \label{eq:results_bosonic_string}
    \tilde{E}_1(R) = \sqrt{\sigma^2 R^2 + 2\pi \sigma (1 - \frac{1}{24})} + \textnormal{const.}
\ee
This relativistic bosonic string theory prediction agrees with our measured first excited energy remarkably
well, see Fig.~\ref{fig:results_first_excited_state}.

\subsection{Gluelumps}
\label{subsec:results_gluelumps}
\begin{figure}[!t]
  \centering
  \mbox{\includegraphics[angle=-90,width=12.0cm]{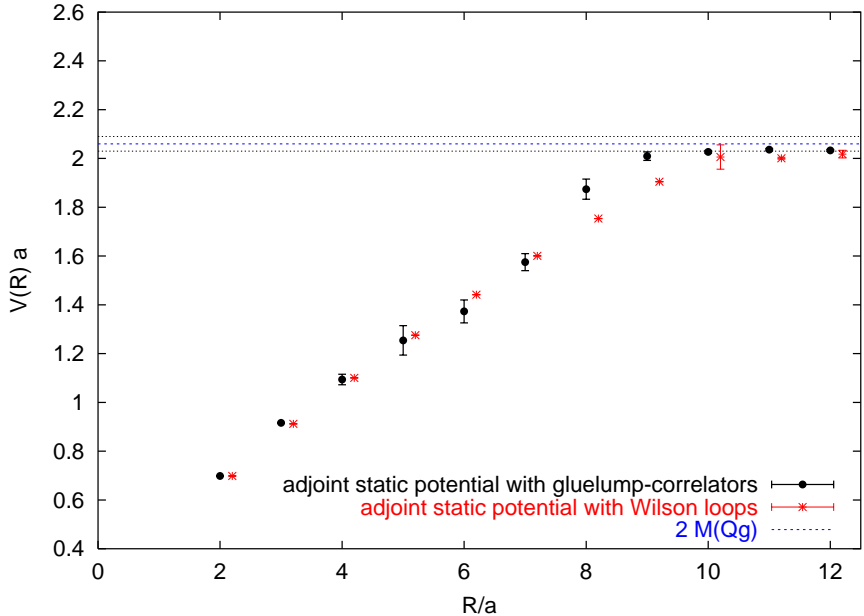}}
  \caption{Agreement of the static adjoint potential $V(R)$ versus $R$,
 extracted from the two-gluelump correlator and the one extracted from Wilson loops only
 (same as in Fig.~\ref{fig:results_stringbreaking_with_fund}, but shifted to the right for clarity).
 The deviations at $R=8a$ and $R=9a$ are due to a large value of the turning point $T_P$, as explained in
 the text.}
  \label{fig:results_gluelump_staticpotential}
\end{figure}
We have seen that the Wilson loop operator has an overlap with both states - the unbroken-string
and the broken-string state. What about the two-gluelump operator?
At large distances $R>R_b$, a single-mass Ansatz is sufficient
since the groundstate is the broken-string state, which has good overlap with
the correlator of two gluelumps, $GG(R,T)$. Therefore, we cannot measure any signal
of the unbroken-string state in this regime.
At distances $R<R_b$, $GG(R,T)$ can be analysed using a two-mass Ansatz
\be \label{eq:results_gluelump_two_mass_ansatz}
     GG(R,T) \sim g_0 e^{-V(R) T} + g_1 e^{-E_1(R) T}\;,\;T > T_{\textnormal{min}}\;.
\ee
At small $R$, the
turning point $T_P$ is small, and the two-mass Ansatz Eq.(\ref{eq:results_gluelump_two_mass_ansatz})
works fine. For $R=8a$ and $R=9a$, which approaches the string breaking distance $R_b$, the energies of
the unbroken-string and broken-string state are almost degenerate, and the turning point value is large.
As mentioned in Subsection \ref{subsec:method_gluelumps}, we used improved spatial transporters to measure Wilson loops.
Here, we use non-improved clovers and have more noise. In addition, we have only one
operator-state and cannot apply a diagonalisation procedure.
Therefore, we have difficulties to measure the subleading groundstate
exponential decay in this regime.
We lose the signal of the unbroken-string state before it becomes visible and cannot extract the unbroken-string
groundstate properly.

Fig.~\ref{fig:results_gluelump_staticpotential} shows the agreement between the static potential
extracted from Wilson loops only and that extracted from the two-gluelump correlator.
This confirms that $GG(R,T)$ has a non-vanishing overlap with the unbroken-string state, similar
to the fact, that the Wilson loop has a non-vanishing overlap with the broken-string state. \\

$GG(R,T)$ can also be used in the multichannel Ansatz as explained in Section \ref{sec:approaches}.
It enters as a diagonal matrix element in the 4x4-matrix $V_{ij}(R,T)$ Eq.(\ref{eq:method_multichannel_matrix}).
We will now consider this approach.

\subsection{Multichannel Ansatz}
\begin{figure}[!t]
  \centering
  \mbox{\includegraphics[angle=-90,width=12.0cm]{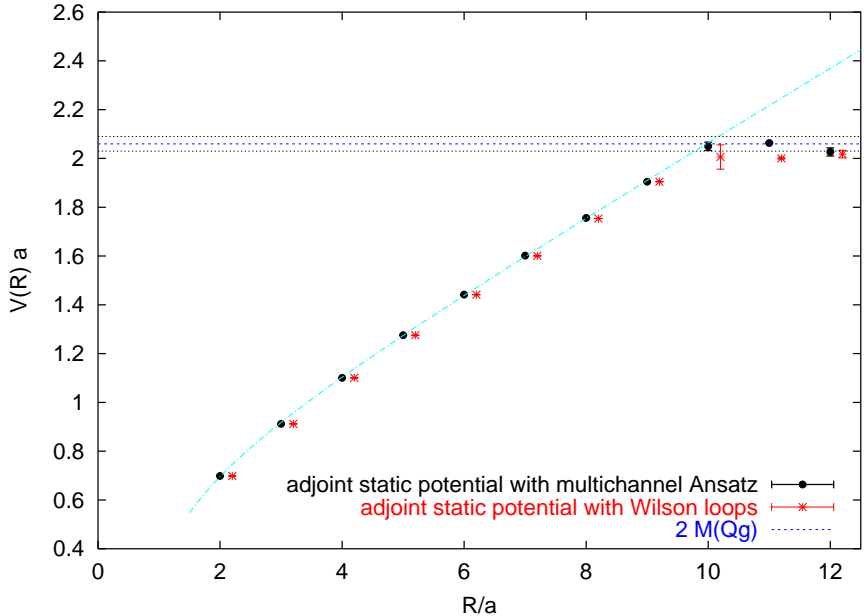}}
  \caption{The static adjoint potentials $V(R)$ versus $R$ using the multichannel
  Ansatz. The agreement with the static potential extracted from Wilson loops only
  (same as in Fig.~\ref{fig:results_stringbreaking_with_fund}, shifted to the right for clarity) is
  good.
The unbroken-string state energy is also drawn (dashed line).
The horizontal line at $2.06(3) a^{-1}$ represents
twice the mass of a gluelump. }
  \label{fig:results_stringbreaking_multichannel}
\end{figure}
The full multichannel matrix Eq.(\ref{eq:method_multichannel_matrix}) is
obtained by including the mixing terms $S_i G(R,T)$ and $G S_j(R,T)$.

\subsubsection{Static potential}

We analyse the $4 \times 4$ matrix using the diagonalisation procedure Eq.(\ref{eq:method_diagonalisation})
at $T_0=4 a$, as described in Subsection \ref{subsec:method_diagonalisation_procedure}.
As expected, a single-mass Ansatz can be applied
for all $R$ using $\lambda^{(0)}(R,T,T_0=4 a)$.
The results, presented in Fig.~\ref{fig:results_stringbreaking_multichannel},
agree with the static potential extracted from
Wilson loops only, with improved accuracy for $R>R_b$. \\

The three Wilson loop operator states at different smearing levels have a good overlap with
the unbroken-string state, but a poor one with the broken-string state. The opposite holds
for the two-gluelump operator state. To confirm this statement, we analyse the overlaps
$a^{(n)}_j(R,T,T_0)$ of Eq.(\ref{eq:method_normalisation}). We consider the
overlap of all three Wilson loop operator states ($j=S_{15},S_{30},S_{60}$) and
the overlap of the two-gluelump operator state ($j=G$) with the groundstate ($n=0$)\footnote{
In $a^{(n=0)}_j(R,T,T_0 = 4 a)$ of Eq.(\ref{eq:method_normalisation}) we fix $T=8 a$. The results are stable
for larger $T$, although with increasing errors.}.
We observe for $a^{(0)}_G(R,T,T_0)$ an abrupt change from $O(10^{-3})$ ($R \leq 10a$) to $O(1)$ ($R>10a$) and
vice versa for $a^{(0)}_S(R,T,T_0)$.
This indicates that string breaking actually occurs at a distance slightly larger than $10a$.

\subsubsection{Excited states}

\begin{figure}[!t]
  \centering
  \mbox{\includegraphics[angle=-90,width=12.0cm]{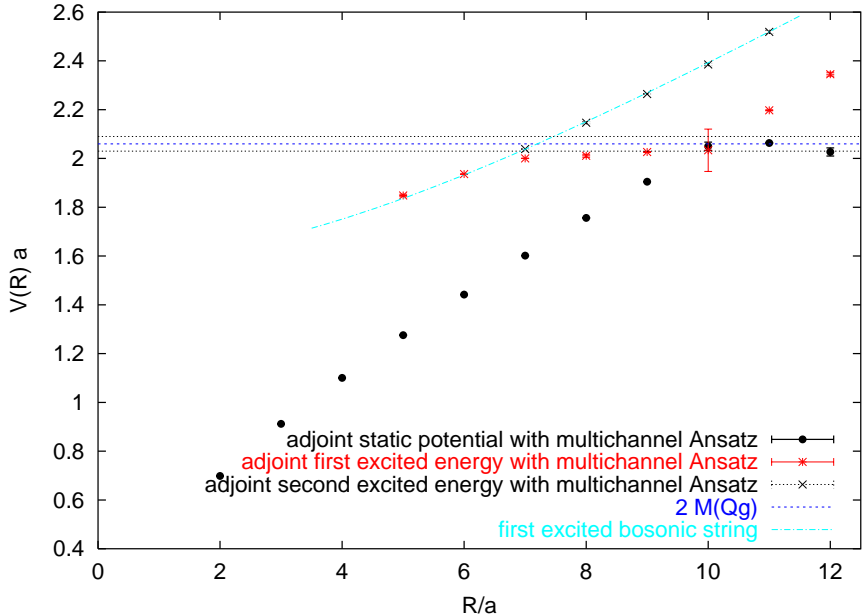}}
  \caption{The static potential $V(R)$, the first and the second excited states energies
  using the multichannel Ansatz.
  We also show the energy $\tilde{E}_1(R)$
  (Eq.(\ref{eq:results_bosonic_string})) resulting from the
  relativistic bosonic string theory.
  The horizontal line at $2.06(3) a^{-1}$ represents twice the mass of a gluelump.}
  \label{fig:results_excited_states_multichannel}
\end{figure}
Considering $\lambda^{(1)}(R,T,T_0=4 a)$, we get information
about excited states by applying a two-mass Ansatz.
For $R > R_b$, we extract the lowest-lying unbroken-string state as expected.
For $R=8 a$ and $R=9 a$, the first excited state is the broken-string state.
For $R=7 a$, the broken-string state and the excited unbroken-string state energies
are almost degenerate. For $R<7 a$ we extract the energy values of the excited unbroken-string
state which are in agreement with the ones extracted using Wilson loops only. For $R<5 a$ we
can no longer extract first excited state energies due to statistical noise.  Finally,
considering $\lambda^{(2)}(R,T,T_0=4 a)$, for $R \geq 7a$, we obtain the energy of the
second excited state, namely the excited unbroken-string state. For $R < 7a$ we
cannot extract the second excited energy due to statistical noise.
In Fig.~\ref{fig:results_excited_states_multichannel}, we show
the static potential extracted from $\lambda^{(0)}(R,T,T_0=4 a)$, the first
excited energy, extracted from $\lambda^{(1)}(R,T,T_0=4 a)$ and the second excited energy,
extracted from $\lambda^{(2)}(R,T,T_0=4 a)$.

\subsection{Polyakov loops}
\begin{figure}[!t]
  \centering
  \mbox{\includegraphics[angle=-90,width=12.0cm]{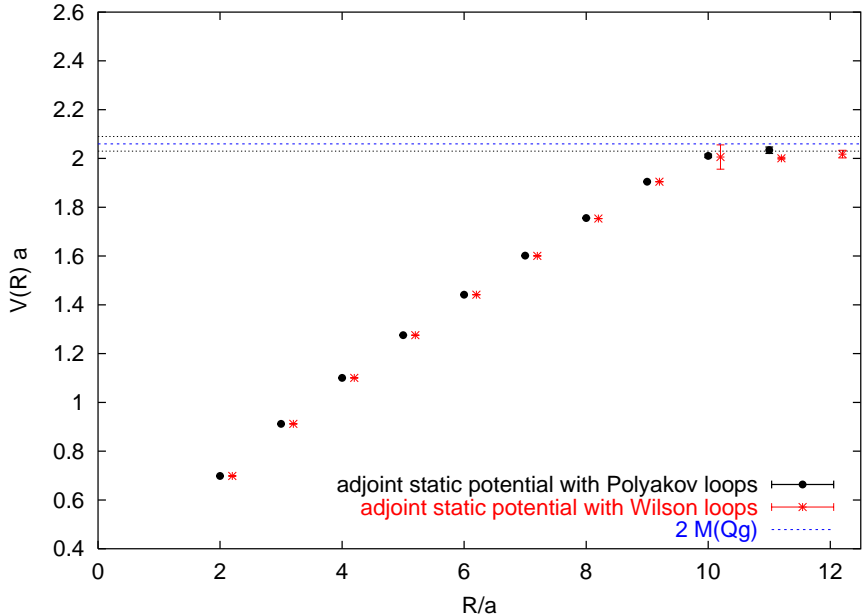}}
  \caption{Polyakov loop method. The static adjoint potential $V(R)$ versus $R$ extracted from the correlator of
  two adjoint Polyakov loops agrees very well with that measured using Wilson loops only
  (same as in Fig.~\ref{fig:results_stringbreaking_with_fund}, shifted to the right for clarity). The flattening
  of the potential can be observed, although we cannot extract the value at $R=12 a$ due to large fluctuations. }
  \label{fig:results_polyakov_loop}
\end{figure}
The correlator of two adjoint Polyakov loops allows the extraction of a temperature-dependent potential $V_{\cal{T}}(R)$
(see Eq.(\ref{eq:approaches_polyakov}))
\be
    V_{\cal{T}}(R) \equiv -\frac{1}{N_t a} \ln \langle P(0) P^*(R) \rangle \;.
\ee
The temperature of our system is ${\cal T} = \frac{1}{N_t a} \approx 30$ MeV
since $N_t=64$ is the temporal extent of the lattice and $a = 0.1022(1)$ fm the lattice spacing.
Since this temperature is quite low, contributions of excited states are negligible. Therefore, we observe
a good match with the static potential measured using Wilson loops
only (see Fig.~\ref{fig:results_polyakov_loop}).

In the regime of
string breaking $R \approx 10 a$ we see flattening of the potential indicating string breaking.
For $R=12 a$, the signal becomes very noisy and the average correlator is negative.

\subsection{Casimir scaling}
\label{subsec:results_casimir}
\begin{figure}[!t]
  \centering
  \mbox{\includegraphics[angle=-90,width=12.0cm]{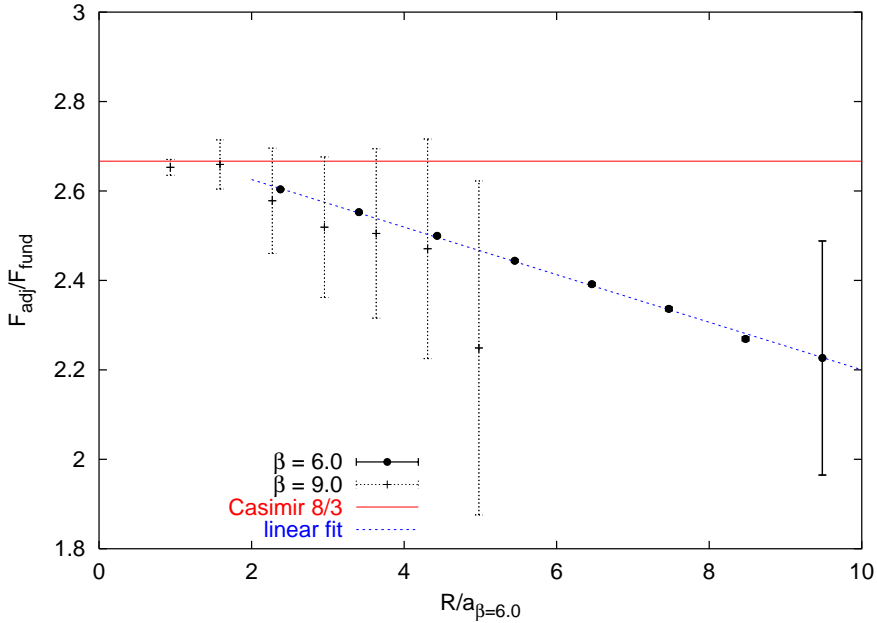}}
  \caption{Ratio of forces $\frac{F_{\textnormal{adj}}(R)}{F_{\textnormal{fund}}(R)}$ as a function of the
  spatial separation. The horizontal line at $\frac{8}{3}$ indicates the Casimir ratio expected
    from perturbation theory. We see clear deviation at distances larger than $R=2 a$. The ratio seems to
    decrease linearly while increasing $R$.}
  \label{fig:results_casimir}
\end{figure}
Since we measure the static potentials between fundamental
and between adjoint charges with high accuracy, we can examine the hypothesis of Casimir scaling,
which says that the ratio of
the adjoint static potential over the fundamental static potential remains equal to the Casimir value $\frac{8}{3}$
over a broad range of distances where both potentials grow more or less linearly with distance.
Already in Fig.~\ref{fig:results_stringbreaking_with_fund}, we see clear deviations from Casimir scaling:
The fundamental static potential, rescaled by the Casimir factor $\frac{8}{3}$, agrees with the adjoint static
potential at small distances $R \leq 2 a$ only. This confirms earlier observations of Ref.~\cite{Poulis:1995nn},
also at $\beta=6.0$, and of Ref.~\cite{Philipsen} at $\beta=9.0$.
Therefore, we apply a more careful analysis
considering forces, defined by
\be \label{eq:results_forceextraction}
    F(r_I) = - \frac{V(r) - V(r-a)}{a}
\ee
where $r_I$ is chosen such that the force evaluated from Eq.(\ref{eq:results_forceextraction}) coincides with
the force in the continuum at tree level \cite{Sommer:1993ce}. The procedure to determine the explicit values
of $r_I$ is described in the Appendix.

In Fig.~\ref{fig:results_casimir}, we show the ratio
$\frac{F_{\textnormal{adj}}(R)}{F_{\textnormal{fund}}(R)}$ at two different
$\beta$'s.
In the regime of perturbation theory, i.e. at small distances $R$, this ratio
is $\frac{8}{3}$ as expected. At larger distances, our $\beta=6.0$ data show clear deviations.
The ratio appears to decrease linearly with increasing distance\footnote{
For $R>R_b$, the adjoint string breaks and the force $F_{\textnormal{adj}}$ is essentially zero. Large fluctuations
at  $R \approx R_b$ induce the large error on the rightmost data point in Fig.~\ref{fig:results_casimir}.}
$R$. The less precise data at $\beta=9.0$ of Ref.~\cite{Philipsen} seem to confirm this
$R$-dependence, making it unlikely to be an artifact of the lattice spacing.

Our results do not necessarily contradict the work of \cite{Bali:2000un}, which found accurate Casimir scaling at
large distances. Ref.~\cite{Bali:2000un} considers the $4d\;SU(3)$ theory, while we consider $SU(2)$ in 3 dimensions.
Rather, what we see might be specific to 3 dimensions. \\

\section{Conclusions}
\label{sec:conclusions}

We have demonstrated that string breaking can be observed, using only Wilson loops
as observables to measure the static potential. This demonstration was performed
in the computationally easiest setup: breaking of the adjoint string in the
$(2+1)$-dimensional $SU(2)$ theory.
Even in this simple case, the unambiguous observation of string breaking, at a
distance $R_b \approx 1$ fm, required the measurement of adjoint Wilson loops
of area in excess of 4 ${\rm fm}^2$, with state-of-the-art variance reduction
techniques. A similar study in $(3+1)$ dimensions with a larger gauge group will
be challenging.

The reason for such large loop sizes is as expected: the Wilson loop has very poor
overlap with the broken-string. Even when the static adjoint charges are separated
by $R > R_b$ and the broken-string becomes the groundstate, its contribution to the
Wilson loop area-law is subdominant. The temporal extent $T$ of the Wilson loop
must be increased beyond a characteristic distance, the turning point $T_P$,
to weaken the unbroken-string state signal and reveal the true groundstate.
We find $T_P \sim 2$ fm, which explains why earlier studies, which did not use
similar variance reduction methods, failed to detect string-breaking.
While large, this turning point value stays well below the strong-coupling
estimate of~\cite{Drummond:1998ar}, which would predict a value about twice as large.

Of course, string-breaking is easy to observe, over a limited Euclidean time extent,
if one uses a multichannel approach where a correlation matrix between unbroken- and
broken-string states is formed and diagonalised, the latter being modelled by a pair
of gluelumps. We reproduce in this case the results in the literature.
We also consider the two-gluelump correlator, which has poor overlap with the
unbroken-string state, and show that the unbroken-string groundstate can be
extracted from that correlator alone, if one allows again for a large Euclidean time
extent. Therefore, full information about the adjoint potential is contained in each of
the diagonal elements of the multichannel matrix.

Finally, we looked in detail at the issue of Casimir scaling, by measuring the
ratio of forces $\frac{F_{\textnormal{adj}}(R)}{F_{\textnormal{fund}}(R)}$ as
a function of $R$. We observe clear deviations of this ratio from the perturbative
value $\frac{8}{3}$, and an apparent linear decrease with $R$. A consistent crosscheck
at a smaller lattice spacing makes this violation of Casimir scaling unlikely to
be a lattice artifact. The situation, however, may be different in the $(3+1)d$ theory.

\section{Acknowledgements}

We are grateful to Michele Pepe and Owe Philipsen for discussions and advice. We also thank
Michele Vettorazzo for helpful hints and support and Oliver Jahn for his continuous help and
for reading the manuscript. The calculations for this work
were performed on the Asgard Beowulf Cluster at the ETH Z\"urich.

\appendix

\section*{Appendix: $r_I$ and the free scalar propagator}
\label{sec:A_free_scalar_propagator}

We follow \cite{Sommer:1993ce} and define forces by
\be \label{eq:A_forceextraction}
    F(r_I) = - \frac{V(r) - V(r-a)}{a}
\ee
where $r_I$ is chosen such that the force evaluated from Eq.(\ref{eq:A_forceextraction}) coincides with
\be
    F(r_I)=C_2 \frac{g_0^2}{2 \pi r_I }\;,
\ee
the force in the continuum at tree level. This results in
\be
     r_I = - \frac{a}{2\pi (G(r,0) - G(r-a,0))}\;,
\ee
where $G(x,y)$ is the massless scalar lattice propagator in 2 dimensions defined by
\be
    - \Delta G(x,y) =\delta(x,y)
\ee
We simply solve for $G(x = n_1 a,y= n_2 a)$ by Fourier transform, namely
\be
    G(n_1 a,n_2 a) = \frac{1}{N^2} \sum_{ l_1=1,l_2=1 }^{N}
         \frac   {     \cos(\frac{2 \pi}{N} \sum_1^2 l_i n_i)  }
                 {     \sum_{i=1}^2 2-2\cos(\frac{2 \pi}{N}  l_i  )} \;,
\ee
where $N \gg n_1,n_2$ is the number of discretisation points, taken sufficiently
large that $G(n_1 a,n_2 a)$ is known to 4-digit accuracy.
Finally, we set the zero-mode contribution so that $G(0,0)=0.0$.
A summary of our results in the range we consider is given in Table \ref{table:appendix_rI}.

\pagebreak

\begin{table}
    \begin{center}
   \mbox{
            \begin{tabular}{|l|l|}
            \hline
                $r/a$   & $G(r,0)$      \\ \hline
0& 0.0000   \\ \hline
1& -0.2500  \\ \hline
2& -0.3634  \\ \hline
3& -0.4303  \\ \hline
4& -0.4770  \\ \hline
5& -0.5129  \\ \hline
6& -0.5421  \\ \hline
7& -0.5668  \\ \hline
8& -0.5881  \\ \hline
9& -0.6069  \\ \hline
10& -0.6237  \\ \hline
11& -0.6389  \\ \hline
        \end{tabular}
        }
               \mbox{
        \begin{tabular}{|l|l|}
            \hline
               $ (r-\frac{a}{2})/a$ & $r_I/a$       \\ \hline
2.5& 2.3790 \\ \hline
3.5& 3.4080 \\ \hline
4.5& 4.4333 \\ \hline
5.5& 5.4505 \\ \hline
6.5& 6.4435 \\ \hline
7.5& 7.4721 \\ \hline
8.5& 8.4657 \\ \hline
9.5& 9.4735 \\ \hline
        \end{tabular}
        }

    \end{center}
    \caption{(left) The scalar lattice propagator $G(r,0)$ in 2d.
    (right) The naive derivation points $\left(r-\frac{a}{2}\right)$ compared with $r_I$ in the range we consider. }
    \label{table:appendix_rI}
\end{table}

\providecommand{\href}[2]{#2}
\end{document}